\documentclass[iop]{emulateapj}     
\usepackage{graphicx} 

\def\vrms{V_\mathrm{rms}}       
\def\lj{\lambda_\mathrm{J}}       
\def\mj{M_\mathrm{J}}             

\def\kms{\,\,\,{\rm km\,s^{-1}}}          
\def\msol{\,{\rm M}_\odot}                
\def\acc{\,{\rm M}_\odot\,{\rm yr}^{-1}}      
\def\mpc2{\,{\rm M}_\odot\,{\rm pc}^{-2}} 
\def\myr{\,{\rm M}_\odot\,{\rm pc}^{-2}{\rm Myr}^{-1}} 
\def\gcc{\,\,\,{\rm g\,cm^{-3}}}   
\def\cm3{\,\,\,{\rm cm^{-3}}}      
\def\kms{\,\,\,{\rm km\,s^{-1}}}   
\def\cc{\,{\rm cm^{-3}}}             
\def\mach{\mathcal{M}}             
\def\machs{\mathcal{M_\star}}       

\def \sigrms{V_{\rm rms}}

\def \nbar{\bar n}             
\def \rbar{\bar \rho}          

\def \ms{\mathcal{M}_\star}
\def \mtil{\widetilde {M}}
\def \rtil{\widetilde {R}}

\def \mjeans{M_J}
\def \fh2{f_{H_{2}}}
\def \vir{\alpha_{vir}}

\def\ga{\,\hbox{\hbox{$ > $}\kern -0.8em \lower 1.0ex\hbox{$\sim$}}\,}
\def\la{\,\hbox{\hbox{$ < $}\kern -0.8em \lower 1.0ex\hbox{$\sim$}}\,}

\begin{document}

\title{Variations of the stellar initial mass function in the progenitors of massive early-type galaxies and in extreme starburst environments}

\author{Gilles Chabrier}
\affil{Ecole Normale Sup\'erieure de Lyon,
CRAL, UMR CNRS 5574, 69364 Lyon Cedex 07,  France,\\
School of Physics, University of Exeter, Exeter, UK EX4 4QL}

\author{Patrick Hennebelle }
\affil{Laboratoire AIM, CEA/IRFU\\
91191 Gif-sur-Yvette Cedex, France }

\and

\author{St\'ephane Charlot}
\affil{UPMC-CNRS, UMR7095, Institut d'Astrophysique de Paris,\\
F-75014 Paris, France }

\date{}

\begin{abstract}
We examine variations of the stellar initial mass function (IMF) in extreme environments 
 within the formalism derived by Hennebelle \& Chabrier.
We focus on conditions encountered in progenitors of massive early type galaxies and starburst regions.
We show that, when applying the concept of turbulent Jeans mass as the characteristic mass for fragmentation in a turbulent medium, 
the peak of the IMF in such environments is shifted towards smaller masses, leading to a bottom-heavy IMF, 
as suggested by various observations. 
In very dense and turbulent environments, we predict that the high-mass tail of the IMF can become even steeper than the standard Salpeter IMF, with a limit for the
power law exponent $\alpha\simeq -2.7$, in agreement with recent observational determinations. This steepening is a 
direct consequence of the high densities and Mach values in such regions but also of the time dependence of the fragmentation process, 
as incorporated in the Hennebelle-Chabrier theory.
We provide analytical parametrizations of these IMFs in such environments,
to be used in galaxy evolution calculations.
We also calculate the star formation rates and the mass-to-light ratios expected under such extreme conditions and show that they agree well with
the values inferred in starburst environments and massive high-redshift galaxies. 
This reinforces the paradigm of star formation as being a universal process, i.e. the direct outcome of gravitationally unstable fluctuations 
in a density field initially generated by large scale shock-dominated turbulence.
 This globally enables us to infer the variations of the stellar IMF and related 
properties for atypical galactic conditions.
 \end{abstract}



\section{Introduction}
\label{intro}

Stars form from the collapse of prestellar dense cores, themselves forming in the overdense regions (clumps) of large molecular gas reservoirs, called giant molecular clouds (GMCs). 
The generic properties of the prestellar core mass function (CMF) and of the resulting stellar initial mass function (IMF)
 are intrinsically associated with the general properties of these clouds.
Various determinations of the IMF in the Milky Way (MW) Galaxy, disk, bulge and nearby star forming regions and young clusters, suggest essentially no or very little variations, with all the inferred IMFs being consistent, within some expected scatter, 
with the same underlying \citet{Chabrier2005} IMF\footnote{The \citet{Chabrier2005} IMF adopts the same form as the \citet{Chabrier2003} one but with a lower normalization at the hydrogen burning limit,
determined from the updated nearby luminosity function released at this time (see \citealt{Chabrier2005} for details). These two IMFs differ essentially in the brown dwarf domain but yield similar M/L ratios in the {\it stellar} regime (see \S\ref{MLR}).} 
 \citep{Chabrier2003,Chabrier2005,Andersen2008,Bastian2010}. 
Similarly, mass-to-light (M/L) ratio determinations in 
spiral galaxies are also consistent with this same IMF \citep[e.g.][]{Portinari2004,vanDokkum2010,Brewer2012,Tortora2013,Tortora2014}.
In contrast, there is now growing evidence from various observations that the IMF in massive elliptical early-type galaxies (ETGs) differs from the previous one, been 
more "bottom heavy", revealing a larger fraction of low-mass stars compared to MW-like environments.
Spectroscopic observations indeed show a marked increase in the strengths of various spectral line absorption features with velocity dispersion in the $\sigma=100$-300 km/s range, pointing 
to the existence of a large population of M dwarf stars and thus a more bottom heavy IMF compared to the MW \citep{vanDokkum2010,vanDokkum2012,Conroy2012a,Conroy2012b,Smith2012,Spiniello2012,Spiniello2014,Goudfrooij2013}.
Furthermore, constraints from observed stellar kinematics and gravitational lensing confirm that massive ETGs have large mass-to-light ratios compared to other galaxies 
\citep[e.g.][]{Treu2010,Thomas2011,Cappellari2012,Cappellari2013,Sonnenfeld2012,Conroy2013,Dutton2013a,Dutton2013b,Barnabe2013,Tortora2013,Tortora2014}. Combined with the former spectroscopic diagnostics, these results suggest 
 that this large mass arises from an unusually large low-mass star population rather than from a population of remnant stars.
An other valuable spectroscopic information is the signature of enhanced $[\alpha/Fe]$ abundance ratios  with increasing large velocity dispersion in the spectra of massive ETGs \citep{Thomas2011,Conroy2012a,Conroy2012b,Conroy2014}, suggesting that the IMF evolves from a Chabrier-like one at abundance ratios close to solar to a more bottom heavy one for highly $\alpha$-enhanced populations. 
Under the conventional assumption that these ratios reflect the star formation timescale, with higher values corresponding to shorter timescales, all these observations suggest that
galaxies harboring $\alpha$-enhanced stellar populations, formed in rapid and intense starbursts, tend to generate a larger population of low-mass stars than galaxies 
with extended star formation histories. The progenitors of massive ETGs, formed in the early universe, are the emblematic representation of such structures. 
 
 The consistent picture that seems to emerge from all these data is that the IMF is not completely "universal" 
but varies from MW-like for spiral galaxies with moderate velocity dispersion formed at low redshift ($z\lesssim 2$), with "quiescent" star formation histories, 
to Salpeter-like 
or even steeper over the entire stellar regime for massive star-forming galaxies, with velocity dispersion $\sigma \gtrsim 200 \kms$, in which stars formed very rapidly, on a "burst" mode, at early epochs. 
As discussed in \S\ref{ETG}, massive ETGs are indeed supposed to be the result of the merging of compact structures formed within large gas inflows in the early universe, 
exhibiting presumably more extreme star formation conditions, gas mean density and velocity dispersion, than the ones prevailing in the MW. 
In the local universe, extreme star forming conditions, resulting from gas rich mergers, prevail also
in starburst environments like the central parts of ultra luminous infrared galaxies (ULIRG) \citep[e.g.][]{Kartaltepe2012}.

In this paper, we explore this issue by examining the dependence of the IMF upon the environment, i.e. the cloud gas
temperature, mean density and turbulent properties,
within the framework of the
\citet{Hennebelle2008,Hennebelle2009,HennebelleChabrier2013} analytical theory of the IMF. This "gravo-turbulent" picture of star formation indeed predicts variations of the IMF with the level of turbulence, in contrast to the standard purely
gravitational star formation scenario, and might naturally explain the aforementioned variations. 
It should be noted that \citet{Hopkins2013} recently explored this issue with a different formalism. Besides the fact that it is important to
verify whether different theories yield similar conclusions, we expand upon Hopkins' results in several ways. First of all by
including the time dependence of the IMF into our formalism, second by calculating the inferred
star formation rates (SFR) and mass-to-light ratios in such extreme environments and third by providing analytical parametrizations of the Chabrier IMF relevant for such conditions.
The paper is organized as follows. We first derive in \S2 relations between the normalizations of Larson relations at cloud scale and those at galactic scale, 
so that the star-forming cloud properties can be inferred directly from the ones of the host galaxy. In \S3, we examine 
the particular conditions prevailing in massive ETG progenitors and how they modify the aforementioned normalizations. 
In \S4, we summarize the Hennebelle-Chabrier theory and highlight its main predictions relevant to the present study.
In \S5, we calculate the IMFs characteristic of extreme star-forming conditions, for which we provide analytical parametrizations.
The SFRs and M/L ratios obtained with these different IMFs are calculated in \S6 and 7, respectively, while section 8 is devoted to the conclusion.


\section{From cloud to galactic properties}
\label{cloud}

\subsection{Larson relations}
\label{larson}

As mentioned above, stars form inside overdense regions, usually denominated "clumps", within giant molecular gas clouds\footnote{Clumps are usually defined as $\sim pc$ size overdense regions in
larger ($\sim 10$-100 pc) molecular clouds and are the very birth sites of star formation. They can be self-gravitating or not and are observed in CO absorption
 or dust emission. They are somehow a scale-down version of GMCs, with slightly smaller sizes and larger mean densities. Besides that, their
global properties are similar to the ones of the clouds so the term "clump" or "cloud" will be used indifferently to denominate the star forming
regions.}. 
We thus consider the properties of the gas in a population of star forming clumps/clouds formed in a given host galaxy.
 In the MW, star forming clumps/clouds are observed to follow the so-called Larson (1981) scaling relations between the cloud H$_2$ mean number density $\nbar$
 or three-dimensional velocity
dispersion $\sigrms$ (which includes both rms and thermal fluctuations), respectively, and its size $L_c$ within a domain ranging from $\sim 1$ to several 100s pc:

\begin{eqnarray}
{\nbar}(L_c)&=&d_0\, ({L_c\over 1\,{\rm pc}})^{-\eta_d}
\label{L1}  \\
\sigrms(L_c)&\equiv& \langle \sigrms^2 \rangle^{1/2}=V_0\, ({L_c\over 1\,{\rm pc}})^\eta.
\label{L2}
\end{eqnarray}
These relations correspond to the nearly equilibrium state of molecular clouds immersed in an ambient medium of constant pressure $P$ (Chi\`eze 1987).
As obvious from these equations, $d_0$ and $V_0$ define the density and velocity dispersion normalization conditions at the 1 pc scale.
Observed values in GMCs in the Galaxy give
$d_0\sim 3\times 10^3\,\cc$ and $V_0\sim 0.8$-1.0$\,\kms$.

For a homogeneous self-gravitating cloud of mass $M_c$ and size $L_c$, the (molecular) gas surface density and pressure read:
\begin{eqnarray}
\Sigma_g&\simeq&{M_c\over \pi L_c^2}\simeq {\rbar}\,L_c /6\label{surden} \\\
 {P\over k_B}&=&\frac{\pi}{2} \frac{G}{k_B}\Sigma_g^2 \simeq 3.34\times 10^3\,({\Sigma_g\over 10\,\mpc2})^2\,\,\,{\rm K\,cm^{-3}} \label{pression}
\end{eqnarray}
where ${\rbar}=\mu \,m_h\nbar$, with $\mu=2.33$ for a cosmic composition and $m_h=1.66\times 10^{-24}$ g, denotes the cloud H$_2$ mean {\it mass} density and
 where we have assumed that essentially all the gas in GMCs is
in molecular form.
 A condition of constant pressure 
thus yields $\eta_d=1$ in eqn.(\ref{L1}). Observations rather suggest $\eta_d\sim 0.7$-1.0 \citep{HennebelleFalgarone2013}, due to the non-homogeneous, arguably fractal nature of GMCs, while the exponent $\eta$ is found to be $\eta\sim 0.4$-0.5.
Strikingly, these scaling relations, illustrating the connection between the properties within a cloud and its ambient medium, are observed to be remarkably universal
 not only under MW-like conditions but in completely different
environments like high redshift star forming galaxies \citep[e.g.][]{Swinbank2011} or 
the MW central molecular zone (CMZ) \citep{Shetty2012,Kruijssen2013}. 
The exponents in these systems remain similar. 
The "universality" of the linewidth-size relation within star-forming GMCs and the fact that it holds over a large range of spatial scales, in particular, suggests that the underlying physical processes driving the ISM dynamics are "universal",
pointing to supersonic turbulence, and provides evidence for large-scale (low spatial frequencies, $> 10$ pc) turbulent driving \citep{Heyer2004,Kritsuk2013}. 
This picture is consistent with GMCs forming on large scales, from colliding flows or global galactic disk instabilities, and inheriting the turbulent properties of these large scale
motions.
 One can indeed relate the value of the exponent $\eta$ to the (three-dimensional) index $n$ of the turbulence velocity power spectrum, 
${\cal P}_{\rm V}(k)\propto k^{-n}$, as \citep[e.g.][]{Hennebelle2008}:

\begin{eqnarray}
\eta={n-3\over 2}.
\label{eta}
\end{eqnarray}
Various simulations of compressible (magneto)turbulence \citep[e.g.][]{Kritsuk2007,Kritsuk2013,Molina2012} 
suggest a value $n\sim 3.8$, which yields $\eta\sim 0.4$, the observed value.

\subsection{Normalization on large scales}
\label{scale}

It is useful to link the dynamical properties of the gas within star-forming clouds, as given by eqn.~(\ref{L2}), the relevant scale for the stellar IMF, 
to the ones of the host galaxy, characterized by a typical length scale $r_{d}$, generally defined as the typical gas scale height $r_{d}\sim h$.
Such a limit between clouds and the surrounding ISM on large scales, however, is rather ill-defined, both theoretically and observationally, and necessarily retains large uncertainties. 
The easiest way to proceed is simply to 
extend the scaling relation (\ref{L2}) up to scales $L_c\sim h$, typical of the largest GMCs, by assuming that the largest turbulence injection scale is $L_i\sim L_c\sim h$.
The velocity dispersion of the star-forming gas in GMCs at the galactic characteristic scale $r\sim h$ is thus the one given by the condition of a Toomre marginally 
stable disk ($Q\sim1$) with gas average volume density $\rho_g$ and mid-plane surface density $\Sigma_0=\Sigma_g(z=0)=2h\rho_g$, 
corresponding to a gas mass  $M_g(r)\approx \pi \Sigma_0(r) r^2$. 
This assumption, which is in substance the one adopted by \citet{Hopkins2012a},
 seems to be generally observationally verified from normal disk to starburst galaxies (e.g. Downes \& Solomon 1998).
The big leap behind this procedure, however, is the assumption that the dynamical properties of the gas in star forming clumps are only marginally affected by
 the transition from the HI to H$_2$ gas or by large scale gradients, so the atomic and molecular gas belong to the same turbulent cascade from galactic scales to cloud scales.
Under such an assumption, the gas velocity dispersion vs size relation for star-forming clumps smoothly decreases from the largest possible turbulent scale, $r=L_i\sim h$, to scales relevant for star 
formation and the IMF, $r\ll h$, while the galactic rotational shear, ${\bar {\omega}} r$, where ${\bar {\omega}} \approx {\sqrt 2}\Omega$ is the disk epicyclic frequency and 
 $\Omega=v_{rot}/r$ its angular frequency, dominates on scales $r\gg h$ \citep[e.g.][]{Hopkins2012a}:

\begin{eqnarray}
\nu^2(r)=\sigrms^2(r)+{\sqrt 3}\,{\bar {\omega}}^2 r^2,
\label{vrms}
\end{eqnarray}
with $\nu(r)\simeq\sigrms(r)$ for $r\ll h$ and $\nu(r)\simeq {\sqrt 3} h\Omega$ for $r\gtrsim h$.
Vertical hydrostatic equilibrium for the gas disk at scale $r\gtrsim h$ implies:
$(\nabla P)_z(r)\approx {\pi G\Sigma_0^2(r)/2h} \approx \rho_g(r)  {\nu^2(r)/ 2h}$.
As mentioned above, this corresponds to the general scaling relation for the velocity dispersion in a marginally stable disk ($Q(r)\sim 1$) on scale $r\sim h$, 
which corresponds to the most unstable scale in a turbulent disk.
This leads to the scaling relation at $r\sim h$ (within geometrical factors of the order of a few):


{\small
\begin{eqnarray}
\nu(r)\approx {\sqrt {2\pi G}}\,\Sigma_0^{1/2}r^{1/2}&\approx& 5.2\,({\Sigma_0\over 10\,\mpc2})^{1/2}({r\over 100\,{\rm pc}})^{1/2} \kms  \nonumber  \\ \\
&\approx& 6.8\,({P/k_B\over 10^4\,{\rm K}\,{\rm cm}^{-3}})^{1/4}({r\over 100\,{\rm pc}})^{1/2}\kms,  \nonumber  \\
\label{normv}
\end{eqnarray}
}
Equating eqn.(\ref{normv}) and eq.~(\ref{L2}) at scale $r\sim L_c\sim h$ provides the normalization condition of eq.~(\ref{L2}), 
which gives a measure of the typical amplitude of turbulent motions at the 1 pc scale, in terms of the one at scale $h$: 
{\small
\begin{eqnarray}
V_0&=&0.16\, ({h\over 100\,{\rm pc}})^{-\eta}\, \nu_h \nonumber \\
&\approx&0.82\,({\Sigma_0\over 10\,\mpc2})^{1/2}({h\over 100\,{\rm pc}})^{0.1} \kms  \nonumber  \\ 
&\approx &1.1  \,({P/k_B\over 10^4\,{\rm K}\,{\rm cm}^{-3}})^{1/4}({h\over 100\,{\rm pc}})^{0.1}\,\,\kms,  
\label{normL2}
\end{eqnarray}
}
where $\nu_h\equiv \nu(h)$ and the numerical factors have been evaluated for $\eta = 0.4$\footnote{We recall that $\vrms$ and $V_0$ denote the 3D velocity dispersion
(${\vrms}_{3D}={\sqrt 3}{\vrms}_{1D}$) and that $\Sigma_0$ denotes the {\it surface} density. Sometimes the {\it projected} density is used instead, which, for a sphere,
introduces a factor 4, as the projected area is $A=\pi(L/2)^2$.}
Note that, in the above estimates, we have assumed for sake of simplicity that the disk consists entirely of molecular gas in GMCs. Considering that these latter only 
contain a fraction of the disk surface density will basically introduce a correcting factor $\fh2\approx f_{GMC}=(\Sigma_{GMC}/\Sigma_0)$ in the above expressions,
changing $\Sigma_0$ by $\Sigma_{GMC}=\fh2\,\Sigma_0$.
 Assuming that most of the galaxy molecular gas is collected into large bound GMCs of size $\sim 100$ pc, $\fh2$ is essentially the galaxy molecular gas fraction.
For MW typical conditions $(\Sigma_0)_{MW}\simeq 20\,\mpc2$, $({P/k_B)_{MW}\simeq 1.5\times 10^4\,{\rm K}\,{\rm cm}^{-3}}$, eqn.(\ref{normL2}) recovers the typical value for $V_0$
mentioned in the previous section.

We can also infer the density normalization. 
Assuming again \citep[e.g.][]{Krumholz2005} that the mass of the most massive star-forming clouds is
about the critical Toomre mass in the galactic disk (which is similar to the two-dimensional Jeans mass for $Q\sim 1$, with the wavelength of the fastest growing mode
 $L_J=\nu^2/\pi G\Sigma_g\sim h$), i.e. $M_c\approx \pi\Sigma_g L_J^2  \approx  \nu^4/\pi G^2\Sigma_g$ (again within numerical/geometrical factors
of the order of a few), one gets, at scale $r\sim L_J\sim h$:

\begin{eqnarray}
{\bar n} &\approx& 10\,\fh2\,({\Sigma_0\over 10\,\mpc2})({r\over 100\,{\rm pc}})^{-1}  \nonumber \\
&\approx& 18\,\fh2\,({P/k_B\over 10^4\,{\rm K}\,{\rm cm}^{-3}})^{1/2}({r\over 100\,{\rm pc}})^{-1}\,\,\cm3,
\end{eqnarray}
and thus, using eqn.(\ref{L1}),
\begin{eqnarray}
d_0 &\approx& 100^{\eta_d}\,({h\over 100\,{\rm pc}})^{\eta_d}\, {\bar n}_h  \nonumber \\
&\approx& 10^3\,\fh2\, ({\Sigma_0\over 10\,\mpc2})  \nonumber \\ 
&\approx& (1.8\times 10^3)\,\fh2\,({P/k_B\over 10^4\,{\rm K}\,{\rm cm}^{-3}})^{1/2}\,\,\cm3,
\label{normn}
\end{eqnarray}
where ${\bar n}_h\equiv{\bar n}(h)$ and where we have used $\eta_d\sim 1$.
This yields a mass-size relation for the GMCs (assuming nearly spherical clouds)

{\small
\begin{eqnarray}
M_c \simeq {\pi \over 6}\,{\rbar} L_c^3 &\approx& (3.0\times 10^5)\,\fh2\,({\Sigma_0\over 10\,\mpc2})({L_c\over 100\,{\rm pc}})^{2}\,\msol \nonumber \\
 &\approx& (5.1\times 10^5)\,\fh2\,\,({P/k_B\over 10^4\,{\rm K}\,{\rm cm}^{-3}})^{1/2}({L_c\over 100\,{\rm pc}})^{2}\,\msol  \nonumber \\
\label{mcloud}
\end{eqnarray}
}
\noindent Equations~(\ref{normL2}) and (\ref{normn}) enable us to relate the normalizations of the cloud mean properties at 1 pc in eqns.~(\ref{L1}) and (\ref{L2}) 
to the ones at galactic scale ($h\sim 100$ pc). In order to avoid the uncertainties in the various numerical coefficients, these relations can be used relative to the values in the MW:

\begin{eqnarray}
{d_{0}\over d_{0_{MW}}} &\approx& \chi_{H_2} {\Sigma_{0}\over \Sigma_{0_{MW}}}          \approx \chi_{H_2}^{1/2} ({P\over P{_{MW}}})^{1/2}\nonumber \\
{V_{0}\over V_{0_{MW}}} &\approx& \chi_{H_2}^{1/2} ({\Sigma_{0}\over \Sigma_{0_{MW}}})^{1/2}   \approx \chi_{H_2}^{1/4} ({P\over P{_{MW}}})^{1/4},
\label{normgal}
\end{eqnarray}
where $ \chi_{H_2}=f_{H_2}/(f_{H_2})_{MW}$ is the ratio of the fraction of molecular gas surface density in the galaxy under consideration compared with the MW value, 
a parameter of order unity.

As mentioned above, these relations implicitely assume that the Larson relations reflecting the
 cloud mean densities and dynamical properties apply all the way through from clump scales to galactic scales, 
regardless of possible large discontinuities either in mean density or turbulence properties
 between star-forming GMCs and the surrounding ISM. 
Such an approximation is probably of questionable validity for nearby galaxies like the MW where
GMCs are discrete, generally gravitationally bound entities much denser than the surrouding ISM (typically $n\sim 100\cc$ vs $\sim 1\cc$) and the ISM is dominated by atomic gas. 
Starburst and high-z massive galaxies, however, 
 are characterized by high surface densities and a larger molecular gas fraction \citep[e.g.][]{Daddi2010}
so that the ISM is a rather continuous star-forming, turbulent medium with a more modest ($\lesssim 10$ or so) density
contrast with the GMCs \citep[e.g.][]{Dekel2009,Ceverino2010}. 
Moreover, the star-forming clumps in starburst environments are confined by a nearly constant bounding pressure during their star formation
history, allowing the use of the Larson relations (Chi\`eze 1987).
Therefore, the above relations should provide reasonable estimates of star-forming cloud conditions in various environments, 
expressed in terms of those typical of the MW.

In conditions such as those leading to the formation of massive ETG progenitors, the external pressure at the surface of the GMC is directly related to
the kinetic pressure of the infalling (circumgalactic) gas, $P\simeq \rho_{inf}v_{inf}^2$.
In a nearly virialized gas disk, the infall velocity is of the order of the virial velocity, $v_{inf}(r) \approx V_{rot}(r)=(GM/r)^{1/2}$ \citep{Genel2012}.
However, although the inflow velocity of cold gas streams is indeed found to be close to the virialization velocity of the isothermal halo for local spiral galaxies
 \citep[e.g.][]{Dekel2009}, this is unlikely to be the case for high-z massive clumpy starburst galaxies such as those leading to the formation of ETG progenitors.
 This lack of equilibrium condition prevents a precise determination of the correlation between gas velocity and virial velocity,
 in contrast to the stellar component, $\Delta V_\star\propto V_{rot}$. 

On the other hand, some (potentially large) fraction $\xi$ of the infall kinetic energy is expected to be converted into turbulent kinetic energy,
leading to a ram pressure, $P_{turb}\sim \rho_g\sigrms^2/k$, at the surface of the star forming clouds \citep[e.g.][]{Bournaud2011,Genel2012}.
The turbulent dissipation rate within the star-forming clouds,
${\dot E}_{dis}\simeq(1/2)\,M_c{\sigrms}^2\ \tau_c$, where $\tau_c={\sqrt 3}L_c/\sigrms$ is the typical crossing time within the cloud, is thus related somehow to the 
gas accretion rate, ${\dot E}_{acc}\simeq 2\,GM_c{\dot M}/L_c$, with some geometrical factor of order unity:

{\footnotesize
\begin{eqnarray}
  {1\over 2}\,{M_c\sigrms^2\over \tau_c} \approx 2\xi\,{GM_c{\dot M}\over L_c} \Rightarrow \sigrms\propto {\dot M}^{1/3} \Rightarrow V_0\propto {\dot M}^{1/3}
\label{rate}
\end{eqnarray}
}
Therefore, at least part of the gas dispersion velocity in the star-forming clouds is expected to correlate directly with the gas inflow rate on the galaxy. 


\section{Massive early type galaxies and extreme starburst environments}
\label{ETG}

In the modern paradigm of structure formation, most of the stars that have ended up today in the cores of massive ($M_\star\gtrsim 10^{10}\,\msol$) ETGs formed in small, 
{\it dense} gas-rich primordial star-forming galaxies, at redshift between $z\sim 3$ and 5, i.e. within a rather short timescale $\Delta t\sim 1$-2 Gyr 
\citep[see e.g.][]{deLucia2007,Wilman2013,Naab2007,Barro2013}.
As mentioned in \S\ref{intro}, measurements of enhanced 
$\alpha$-element abundances in the spectra of these galaxies indeed suggest a star formation timescale of a few Myr \citep{Thomas2005,Conroy2014}.
Hence, star formation in the progenitors of massive ETGs presumably occured more in burst modes than in gradual "disk" modes typical of star-forming galaxies at lower redshift.
Then, these primordial structures continued to grow mainly through dry (gas-poor), minor merger events, to eventually lead to the formation of the massive ETGs observed today,
 without producing significant new stars (see references above).
The early phases of star formation in massive ETG progenitors thus involve violent dynamical
dissipational processess such as gas-rich mergers or dynamical instabilities fed by strong infalling gas flows from the IGM, that induce compact starbursts in the central regions. 
Gas consumption into stars in the central starburst, ram pressure on dust 
grains or outflows driven by stellar or AGN feedback mechanisms then quench star formation on very short (dynamical) timescales and the galaxy then evolves passively.
Therefore, according to this paradigm, star formation in massive ETG progenitors occured thoughout bursts in dense/compact galaxies and was driven primarily by
intense cold gas accretion events, originating from gas streams along the filaments of the cosmic web, efficiently forming stars centrally rather than in an extended disk 
\citep[e.g.][]{Genel2012}.
The IMF was imprinted during these early phases of intense gas-rich accretion and remained barely affected afterwards.
Conditions quite different from the ones prevailing in quiescent MW-like spiral galaxies.


\subsection{Density }

As just mentioned, the high-redshift ($z\gtrsim 2$) progenitors of massive ETGs are much more compact, with sizes smaller by a factor $\sim 6$, than bulges or spiral galaxies of similar stellar mass in the local universe \citep{Barro2013,Williams2014}. 
The combination of high masses and small sizes results in extremely high densities. Similar conditions are expected in the central regions of massive elliptical galaxies that are thought to form in compact starbursts \citep[e.g.][]{Diamond-Stanic2012}.
The average gas surface densities in star forming clouds in starburst galaxies range from $\Sigma_g\sim 10^3$ to $10^5\,\mpc2$ \citep{Turner2000,vanDokkumEA2008}, 
compared to a typical range $\sim 1$ to $10^2\,\mpc2$ within a typical Milky Way disk GMC (Kennicutt 1998, Bryant \& Scoville 1999).
After eqn.(\ref{pression}), such high densities correspond to interstellar ISM pressures as high as $P/k_B \gtrsim \,10^{7}$-$10^{11}$ K cm$^{-3}$,
compared to $P/k_B \sim \,10^{4}$ K cm$^{-3}$ for the typical MW ISM pressure. 
  
According to the relations derived in \S\ref{scale}, such a dense and highly pressurized ISM  
leads to the formation of massive (up to $\sim 10^{10}\,\msol$) and dense  star-forming regions, a factor $\gtrsim 10^4\times$ higher than the largest characteristic mass of GMCs in the MW \citep[e.g.][]{Elmegreen2007,Swinbank2012}.
Using eqns.~(\ref{normn}) and (\ref{normgal}), the density normalization at 1 pc is expected to reach values as high as $d_0\sim 10^6\cc$ in such starburst regions.
This means that, under such
high-pressure conditions, the molecular gas can be more than $\sim1000\times$ denser 
{\it at all scales} than the gaseous star-forming clumps in the MW.
Indeed, the compact central starbursts and central regions of ULIRG-like galaxies (gas-rich mergers) have characteristic gas densities that are $10^2$ to $10^4$ times higher than the average for normal disks, 
with the dominant gas mass fraction lying at densities $n$(H$_2)\gtrsim 10^5\cc$ \citep{Gao2004,Diamond-Stanic2012}.
Extreme starburst regions like Arp 220 are larger than ordinary GMCs, but are filled with molecular gas at a density usually found only in small cloud cores (Downes \& Solomon 1998).

\subsection{Velocity dispersion}

As mentioned above, the most admitted explanation for the rapid and significant early mass growth of the progenitors of the most massive ETGs involve violent dynamical processes
 such as repeated gas-rich ("wet") mergers 
or strong accreting gas flows at high redshift, with most of their stellar mass being
 formed by $z\sim$ 3.
Accretion rates onto the galaxy from the intergalactic medium reach values as large as ${\dot M}=10^2-10^3\,\acc$, to be compared with ${\dot M}\approx2\,\acc$ for the MW
 \citep[e.g.][]{Dekel2009,Ceverino2010,Klessen2010}.
These accreting flows
produce a highly turbulent velocity field
that pervades the whole region as the gas is entrained into the flow.
Globally, the combined effects of large accretion rates, larger densities and higher pressures relative to the local ISM
all yield a systematic enhancement of turbulent velocities {\it at all scales}. 
According to eqn.~(\ref{normL2}) and (\ref{rate}), we expect a factor $\sim 5$-10 increase in the normalization factor $V_0$ at the 1 pc scale, yielding values as
high as  $V_0\sim 10\,\kms$ under the aforementioned conditions.
The CMZ, for instance, characterized by larger ambient pressure and larger densities and temperatures than the local ISM, 
is found to exhibit velocity dispersions systematically higher, 
resulting in a scaling coefficient 
approximately $5\times$ larger than for GMCs in quiescent environments like the standard MW clouds (e.g. \citealt{Swinbank2011} Fig. 6, \citealt{Shetty2012} Fig. 8). 
And high-redshift galaxies are indeed characterized by a considerably higher degree of internal turbulence than present-day galaxies of comparable mass \citep{Genzel2008}.

\subsection{Temperature}
\label{temp}

The temperature of the gas in a molecular cloud depends on many different physical processes.
To the best of our knowledge, detailed calculations have not been conducted for ETG progenitors or starburst environments so it is difficult to infer the typical gas temperature under such conditions. 
This latter, however, is intuitively expected to be significantly larger than the typical $T\sim 10$ K value
representative of relatively quiescent, low-density star-forming molecular clouds.
First of all, as mentioned above, star formation in the progenitors of present ETGs is expected to have occured through an initial burst at $z\sim$ 5, yielding a minimum background temperature of about $\sim 20$ K.
Second of all, the predominance of gas compression due to gas inflows and
shock dominated turbulent motions yields large compressional heating rate ($\propto {\sqrt \rho}$).
 At first order, the dissipation of this kinetic energy into thermal energy will raise the gas temperature, even though part of this energy will be
radiated away.
In extreme starbursts like those in low-redshift ULIRGs, the infrared emission peaking at wavelengths $\sim$60-100 $\mu$m reveals dust heated at temperatures 
$T_d\sim 60$-70 K in molecular clouds \citep{Downes1998,Greeve2009}. 
At the high densities typical of the regions of interest, $\nbar \gtrsim10^4\gcc$, gas-dust energy exchange becomes quite efficient, so the kinetic temperature of the (thermally coupled) gas reaches similar values.
Of course, what matters is the temperature of the gas before star formation sets in, so the comparison with ULIRGs might be questionable. The consequences of the uncertainties in the gas temperature on
the IMF will be addressed in \S\ref{IMFvar}.

In the calculatons below, we will examine the impact of such high density, pressure, temperature and velocity dispersion conditions upon the IMF. 
A word of caution, however, is necessary. Despite the support provided by the above general analysis, our assumption that 
the physical conditions in ETG progenitors and in starburst environments are similar could be questioned.
The temperature determination, for example, is particularly uncertain. The vicinity of an HII region, common in spiral galaxies as opposed to ellipticals, or of a black hole, for instance, might heat up the gas significantly, 
thereby decreasing the cloud typical Mach number (${\cal M}\propto T^{-1/2}$). Dense, high-pressure environments thus do not necessarily imply highly turbulent conditions. These uncertainties must be borne in mind when trying to characterize ETG progenitor conditions.


\section{Hennebelle-Chabrier theory. General formalism and main features.}
\label{HC}

Recently, \citet[][HC08]{Hennebelle2008} have developed an analytical theory of the IMF, extending the Press \& Schechter formalism developed in cosmology for
linear density fluctuations to the gravoturbulent picture of star formation, characterized by highly non-linear fluctuations. 
This theory successfully reproduces within the same framework the observed distribution of unbound CO-clumps and of gravitationally bound prestellar cores\footnote{As discussed below, the HC theory is essentially devoted to the formation of gravitationally bound prestellar cores
 and thus is truncated at scales larger
than the cloud scale. Within this limit, the theory correctly accounts for the cloud in cloud problem in the original Press-Schechter formalism \citep[see HC08][]{Hopkins2012b}.}. 
Both the clump and the core mass functions only depend on one single parameter, namely the index of the log density and velocity power spectra 
 of turbulence. 
This index is indeed found to be similar for both spectra in simulations of shock dominated turbulence, in the explored range of Mach values, 
with a characteristic (3D) value $n^\prime\simeq n\simeq 3.8$ \citep[e.g.][]{Kritsuk2007}, between the Kolmogorov and Burgers values. 
In the HC formalism, the normalization and the width of the IMF are then entirely determined by the characteristic conditions of the parent molecular cloud,
 mean density ${\bar n}$, temperature $T$ and large scale Mach number ${\mathcal M}=\sigrms/C_S$, where $C_S\simeq 0.19\,(T/10\,{\rm K})^{1/2}(\mu/2.33)^{-1/2}\kms$ is the sound velocity.

In a subsequent paper, \citet[][HC09]{Hennebelle2009} took into account the thermodynamic of the gas and showed that it has a significant impact on the low-mass part of the spectrum.
More recently, the same authors have included the time-dependence 
in their theory, extending the calculations to an analytical derivation of the star formation rate 
\citep[][HC11, HC13]{Hennebelle2011a,HennebelleChabrier2013}. 
This leads to some modifications of the low-mass part of the IMF compared with the static theory and predicts star formation rates in very good agreement with the observed values in Milky Way molecular clouds (see HC13).

Although alternative IMF theories have been suggested \citep[see e.g.][for recent reviews]{Hennebelle2011b,Offner2014}, 
the most achieved one, besides HC, is the one recently derived by \citet{Hopkins2012a}, based on the so-called excursion set formalism also used in cosmology.
Both the HC and Hopkins theories rely on the concept of structure (clouds or cores) formation being induced by density fluctuations, $\delta=\ln(\rho/\rbar)$,
induced by the small scale dissipation of large scale supersonic turbulence. 
The random field of density fluctuations is thus given by the PDF (i.e. power spectrum) of turbulence. 
For isothermal (magnetized or non-magnetized) turbulence, this latter has been found in many studies to be well reproduced by a lognormal form \citep[see e.g.][and references therein]{Vazquez-Semadeni1994,Federrath2013}, i.e. a Gaussian field in logarithm of the density, characterized by a 
variance $\sigma(\delta)$. 
We stress, however, that both HC and Hopkins theories remain valid for any choice of the PDF, even though a lognormal form greatly simplifies the calculations.
Smoothed at scale $R$, the random field of fluctuations $\delta_R\equiv\delta(R)=\log(\rho(R)/\rbar)$ is thus given by:

\begin{eqnarray}
{\cal P}_R(\delta_R)&=&{1\over \sqrt {2\pi\sigma(R)^2}}\exp\Big\{-{\delta_R+{\sigma(R)^2/ 2}\over 2\sigma(R)^2}\Big\}\\
\sigma(R)^2&=&\int_{2\pi/L_c}^{2\pi/R}\delta^2(k)W_k^2(R)d^3k=\sigma_0^2\Big[1-({R\over L_c})^{n^\prime-3}\Big] \nonumber \\
\label{sigma}
\end{eqnarray}
Here $W_k$ is a window function, chosen to be the sharp-$k$ space truncated function.
Various simulations of supersonic turbulence (see above references) yield for the variance of the unsmoothed density field, $\sigma_0$:

\begin{eqnarray}
\sigma_0^2=\ln\,[1+(b\mach)^2],
\label{sigma0}
\end{eqnarray}
where $b$ describes the relative importance of the compressible and solenoidal contributions to turbulence forcing, with $b\approx 0.3$ and $1$ for purely solenoidal and compressive
 modes, respectively, and $b\approx 0.4$ for equipartition between the modes \citep{Federrath2010}. 
The large-scale Mach number $\mach\equiv \mach(L_c)=\sigrms(L_c)/C_s$ itself obeys the scaling relations (\ref{L2}) and (\ref{normL2}). 

One of the differences between the HC and Hopkins theories is that in the first one, the scale dependence (eqn.~(\ref{sigma})) derives from the turbulent log-density power spectrum, supposed to obey a power law of index $n^\prime$ whose value,
as mentioned above, is found in simulations to be similar to the one of the {\it velocity} power spectrum $n$, with $n^\prime\approx n\approx 3.8$ (Kritsuk et al. 2007). 
Although other forms of scale dependence are certainly possible, this one seems to be reasonable, as it relies on the properties of compressible turbulence inferred in simulations and seems to be corroborated by studies aimed at exploring this issue \citep{Hennebelle2007,Schmidt2010}. 
In contrast, in Hopkins' theory, the Mach dependence of the variance $\sigma(R)$ of the PDF is simply given by the assumption that eqn.~(\ref{sigma0}) applies at all scales, from cores to
galactic scales. Although also plausible, at least for an isothermal gas, 
this assumption, however, remains to be verified. Indeed, it is not clear whether such a condition, which intrinsically implies 
that the density distribution smoothed on a given scale only depends on the gas properties at that scale and not at larger scales, 
adequately represents the frontier between star forming molecular clumps and the ISM, and whether the relation still holds for compact, clumpy galaxies.
The other difference between the two theories concerns the divergence of the integral in eqn.~(\ref{sigma}) on large scales. In both formalisms, the size of the largest turbulence-induced fluctuation is the turbulence outer injection scale itself. As
discussed in \S\ref{scale}, the maximum value for this latter is typically the galaxy scale height $h$. This in turn sets up the maximum size of a fluctuation in eqn.~(\ref{sigma}),
 $R_{max}\sim L_c\sim h$. 
It is clear that, when approaching the turbulence injection scale, both the overdensities and the variance of the fluctuations must vanish.
In the HC formalism, the power spectrum is simply truncated at these scales according to eqn.~(\ref{sigma}). As acknowledged in HC08, the HC theory thus becomes of dubious validity at large
scales ($R\sim L_c$).
Hopkins provides a solution to avoid this divergence by noting that
on large scales, i.e. when the scale approaches or becomes typically larger than the galactic scale height, $R\gtrsim h$, the gas velocity becomes
dominated by the shear velocity, ${\bar {\omega}} R$, which becomes responsible for damping of the density fluctuations (see \S\ref{scale}). 
Assuming, as mentioned above, that relation (\ref{sigma0}) generalises on a scale-by-scale basis on all scales, from $R\ll h$ to $R\gtrsim h$, 
this provides a natural, although no longer analytic truncation of $\sigma(R)$ in Hopkins' theory. 

 The above uncertainties in both theories when switching from cloud scales to galactic ones illustrate the rather ill-defined border between clouds and the ISM. 
In the present paper, we will use the same HC formalism simply by assuming that cloud sizes in eqn.~(\ref{sigma}) extend up to $L_c\sim h$, typical of the size of the largest GMCs, 
with the normalization conditions for the clouds being related to the galactic properties by the relations derived in \S\ref{scale}.
As mentioned above, in principle the HC theory becomes dubious at such large scales. The present study, however, is devoted to the {\it stellar} (or prestellar core) mass function, 
i.e. to the formation of stars within giant molecular clouds, {\it not} to the formation or mass function of the clouds themselves. 
When applied to scales relevant for prestellar core formation ($R\sim 0.1\,{\rm pc}\ll L_c$), the HC formalism has been shown to successfully 
 reproduce the observed CMF of both bound prestellar cores and unbound overdense CO clumps (HC08, HC09). 
 
In the HC formalism, fluctuations of size $R$ and mass $M_R=M(R)$ which are prone to collapse, leading to the formation of self-gravitating prestellar cores,
  are the ones exceeding a density threshold, $\delta_R\ge \delta_R^c$, 
given by the virial condition (see HC08). In Hopkins' excursion set formalism, $\delta_R^c$ represents the density "barrier". Note that,
in both formalisms, the threshold or barrier depends on the scale $R$, in contrast to the cosmological case.
The mass within a fluctuation of scale $R$ is $M_R=C_m\rho_RR^3$, where $C_mR^3$ and $\rho_R$ are the associated typical volume and mean gas mass density,
while $C_m$ is a spatial {\it filtering} factor of the random density field, whose value depends on the window function and can vary by a factor of a few (Lacey \& Cole 1994).
The masses and sizes of the fluctuations will be writen in units of the Jeans mass/length, $\mtil(R)=M(R)/\mj$, $\rtil=R/\lj$ with
{\small
\begin{eqnarray}
\mj&=& a_J\,{ C_s^3 \over \sqrt{G^3 \bar{\rho}}}=C_m \rbar \lj^3 \nonumber \\
&\approx& 0.8\,\, ({a_J\over C_m}) \,({T \over 10\,{\rm K}})^{3/2}\,
({\mu \over 2.33})^{-2}\,({{\bar n} \over 10^4\,{\rm cm}^{-3}})^{-1/2}\, \msol 
\label{mjeans} \\
\lj&=& \left( {a_J \over C_m} \right)^{1/3}{C_s\over \sqrt {G{\bar \rho}}} \nonumber \\
&\approx& 0.1\,\left( {a_J \over C_m} \right)^{1/3} \,({T \over 10\,{\rm K}})^{1/2}\,
({\mu \over 2.33})^{-1}\,({{\bar n} 
\over 10^4\,{\rm cm}^{-3}})^{-1/2}\,\, {\rm pc},\nonumber \\
\label{ljeans}
\end{eqnarray}
}
where $\mu$ is the mean molecular weight ($=2.33$ for cosmic composition), $T$ the temperature of the gas and $a_J$ is a {\it geometrical} factor. For a uniform sphere, $a_J={\pi^{5/2}/ 6}$, and a top hat function in the real space, $C_m=\pi/6$, one gets the standard expression,
$\lj  ={\sqrt \pi } \,{C_S\over (G\rbar)^{1/2} }$, with the Jeans mass being the mass enclosed in a sphere of diameter $\lj$. 
Because of the uncertainties on the smoothing filter (the window function), be it theoretical, numerical or observational,
 and the shape/structure of the fluctuations, these masses and sizes
inevitably retain some degree of uncertainty. If the star forming clumps are filamentary, for instance, the mean thermal Jeans mass will differ by a factor $\sim 0.6$ from that of a sphere \citep{Larson2003}.
Given this uncertainty, we will simply adopt in the present paper for the Jeans length: $\lj\simeq {C_s/ \sqrt {G{\bar \rho}}}$. 
Adopting a definition with a prefactor different from 1 simply translates into a uniform shift of the calculated IMF in mass, and is thus degenerate with the value of the CMF-to-IMF (core-to-star) mass conversion efficiency (see below). 
All these factors remain of the order of unity.

In both the Hennebelle-Chabrier and Hopkins theories, turbulence plays a key role in yielding the proper CMF and in determining the Salpeter-like slope at high masses \citep{Chabrier2011}, as confirmed by numerical simulations \citep{Schmidt2010}. 
A fundamental outcome of these theories is the concept of "turbulent Jeans mass", which is the mass at a given scale $R$ which fulfills the aforementioned 
threshold condition for gravitational collapse in a {\it turbulent} density field. In the HC formalism, this mass reads:

\begin{eqnarray}
\mtil (\rtil)={M(R)\over \mjeans}=\rtil \,(1+\ms^2\rtil^{2\eta}),
\label{mcrit}
\end{eqnarray}
where the parameter $\machs$ measures the importance of turbulence at the {\it Jeans} scale $\lj$, as opposed to the Mach number $\mach$ at the {\it cloud} scale $L_c$  (see HC08):

\begin{eqnarray}
\mach &=& {V_0\over C_s}({L_c\over 1\,{\rm pc}})^\eta = {\cal M}_h\, ({L_c\over h})^\eta\\
\machs &=& {1\over \sqrt 3}{V_0\over C_s}({\lj\over 1\,{\rm pc}})^\eta  = {{\cal M}_h\over \sqrt 3}\, ({\lj\over h})^\eta,
\label{mach}
\end{eqnarray}
where ${\cal M}_h=\nu(h)/C_s$ denotes the Mach value at scale $h$ and where we have made use of the scaling relations derived in \S\ref{scale}. 
As explained in detail in \citet{Chabrier2011}, the role of turbulence in defining such a characteristic mass should not be considered in a static (pressure like) sense, since turbulence has already dissipated by the time the prestellar core is formed. 
But rather in a statistical or dynamical sense, in selecting in the very initial field of density fluctuations those dense enough to not be dispersed by the flow before they have a chance to collapse. 
What matters in turbulence is thus the rms {\it velocity} rather than pressure. 
Eqn.~(\ref{mcrit}) determines the transition between the thermally dominated ($\machs\rightarrow 0$) and the turbulent dominated 
($\ms^2R^{2\eta}\gg 1$) regimes, with the respective scaling relations for collapsing structures:

\begin{eqnarray}
{\rm thermal} &:&  M(R) \propto R,\,\,\,\rho(R)\propto R^{-2} \label{mrther}\\
{\rm turbulent} &:&  M(R) \propto R^{n -2},\,\,\,\rho(R)\propto R^{n-5},
\label{mr}
\end{eqnarray}
where we have used eqn.~(\ref{eta}). We recall that $n\sim 3.8$ denotes the three dimensional velocity power spectrum
index of turbulence. Note that for $n= 4$ (Burgers pressureless regime), i.e. $\eta=1/2$, we recover exactly the scaling relations (\ref{normv}) for
the rms velocity dispersion of star-forming clouds at constant pressure.
It is worth mentioning that relation~(\ref{mr}) yields for the bound prestellar cores $M \propto R^{1.8}$, consistent with the
observational determination from Herschel, $M \propto R^{\beta}$, with $\beta\sim 1$-2 \citep{Andre2010,Konyves2010}.
 
The transition limit between thermally dominated and turbulence dominated regimes defines the equivalent of a sonic scale and a sonic mass:

{\footnotesize
\begin{eqnarray}
\rtil_s \simeq \ms^{-1/\eta}\,&\Rightarrow&\, R_s\simeq 3^{1/2\eta}\,({V_0\over C_s})^{-1/\eta}\,{\rm pc}\simeq  4.0\,{\cal M}_h^{-1/\eta}\,h \label{rads}  \\ 
\mtil_s\simeq 2\,\rtil_s &\Rightarrow&\, M_s \simeq 2\,\machs^{-1/\eta}\,M_J.  \label{rs}
\end{eqnarray}
}
The numerical coefficient in eqn.~(\ref{rads}) has been evaluated for $n=3.8$, i.e. $ \eta=0.4$. 
These values are similar to those found in \citet{Hopkins2012a}.
Equation~(\ref{rads}) indicates that, in the HC formalism, the sonic length is entirely determined by the Jeans scale and the level of turbulence (Mach number) at this scale, 
illustrating the respective roles of gravity and turbulence in the mass-size relation of collapsing cores.
Under typical MW conditions ($V_0\simeq 0.8\,\kms$, $C_s\simeq 0.2\,\kms$, then $\machs\sim \sqrt 2$), this corresponds to $R_s\sim \lj$, $M_s\sim M_J$.

In the time-dependent version of the HC theory \citep{Hennebelle2011a,HennebelleChabrier2013}, the number density mass spectrum of gravitationally bound cores reads:

{\small
\begin{eqnarray}
{\cal N} (\widetilde{M}_R ) &=& {dn\over d\mtil_R} \nonumber \\  &=&
{\scriptscriptstyle 
 {2\ \over \phi_{t}}  {\cal N}_0 \, { 1 \over \widetilde{R}^6} \,
{ 1 + (1 - \eta){\cal M}^2_* \widetilde{R}^{2 \eta} \over
[1 + (2 \eta + 1) {\cal M}^2_* \widetilde{R}^{2 \eta}] }
\times    \left( {\widetilde{M}_R \over \widetilde{R}^3}  \right) ^{-1 -   {1 \over 2 \sigma^2} \ln (\widetilde{M}_R / \widetilde{R}^3) }
\times {\exp( -\sigma^2/8 ) \over \sqrt{2 \pi}\, \sigma } },\nonumber \\ 
\label{grav_tot2}
\end{eqnarray}
}
\normalsize

where ${\cal N}_0=  \bar{\rho} / M_J$ and $\phi_t$ is a dimensionless timescale factor which determines the typical time $\tau(R)$ within which a new density fluctuation of scale $R$ 
is generated in the density field after the former one has collapsed, $\tau(R)=\phi_t \tau_{ff}(R)$, where $\tau_{ff}(R)$ denotes the fluctuation free-fall timescale. 
Theoretical estimates \citep{Krumholz2005,Hennebelle2011a} and numerical simulations \citep{Federrath2012} suggest $\phi_t\approx 2$.
Strictly speaking, the mass spectrum in the HC theory entails a second member, which can also be calculated analytically (see Appendix B of HC08).
However, as shown in HC08, this term becomes significant only when $R\sim L_c$ and will only impact the highest mass part of the IMF. 
For sake of simplicity, this term will be dropped in the present calculations.

It should be noticed that the theory yields the mass spectrum of {\it prestellar} cores, i.e. the CMF, not the final IMF. 
As mentioned in \S6, there is observational evidence that the latter one strikingly resembles the former one, with a uniform core-to-star mass conversion efficiency $\epsilon =M_\star/M_{core}\sim 0.3$-0.7, 
due to the magneto-centrifugational outflows associated with the birth of the protostar \citep{Matzner2000}. Since, as noted above, the ambiguity on the precise
value of the geometrical and filtering parameters and thus on the Jeans length also translate into a uniform shift of the CMF, there is clearly a degeneracy between
these two factors. This uncertainty, however, does not affect the general purpose and conclusions of the present study.
Therefore, for sake of simplicity, we will simply assume that the CMF (\ref{grav_tot2}) represents the IMF.
 As seen from the above equations, the IMF only depends on the cloud's mean density $\bar{\rho}$ for the normalization and the 
variance $\sigma^2$ of the turbulence PDF for the shape (width and peak). 
As shown in eqn.~(\ref{sigma}), this latter quantity only depends on the index $n^\prime$ of the power spectrum of the log(density) of turbulence $P_{\ln\rho}(k)\propto k^{n^\prime}$, with $n^\prime\sim n\sim 3.8$.

As demonstrated in HC13, the time dependence affects the static mass spectrum quantitatively through the normalization factor $1/\phi_t$ but also qualitatively through 
the modification of the exponent,
$-1 -{1 \over 2 \sigma^2} \ln (\widetilde{M}_R / \widetilde{R}^3)$, instead of $-3/2 -   
{1 \over 2 \sigma^2} \ln (\widetilde{M}_R / \widetilde{R}^3)$ in the static theory, which arises from the time dependence of the collapsing cores, proportional to $\sqrt{\rho(R)} \propto (M_R/R^3)^{1/2}$ (cf HC13).
This yields a steepening of the high-mass slope of the IMF\footnote{It must be kept in mind that $ \widetilde{R}$ depends on $ \widetilde{M}$ (cf. eq.(\ref{mcrit}) above). This makes the IMF steeper in the time-dependent case (see eqns.(24) and (25) of HC13).}. 
Physically speaking, this stems essentially from the fact that, during the collapse of the cloud, high-mass cores have time to fragment into smaller ones, an effect not accounted for in a static theory of the IMF. Time-dependence also promotes the number of small cores, because of their shorter free-fall timescale, $\tau_{ff}(R)\propto 1/\sqrt{\rho(R) }$, for a density fluctuation $\rho(R)$.
As examined below, this steepening of the IMF bears important consequences in very dense and turbulent environments.

As seen from (\ref{grav_tot2}) (see also HC08 \S5.4), the mass function 
entails a lognormal and a power-law contributions. This latter one is dominant in the mass regime:

\begin{eqnarray}
{\rm power}\,{\rm law}:\,\,\,\,\, e^{-2\,\sigma^2} \ll \widetilde{M} \ll e^{+2\,\sigma^2}, 
\label{lognorm}
\end{eqnarray}
determined by the variance of the PDF, while the lognormal part produces an exponential cut-off outside these limits at small and large masses. 
Therefore, the stronger the turbulence (the higher the Mach) the larger the mass range covered by the power law part of the IMF 
and the smaller the mass at which it turns over a lognormal form.
 In the high-mass, turbulence dominated regime ($\machs^2 \rtil^{2\eta}\gg1$),
 the power law part reads ${\cal N}(\mtil)\propto \mtil^{-\alpha}$, with $\alpha=\alpha_1+\alpha_2$ and

\begin{eqnarray}
\alpha_1& = &{4+2\eta \over 2 \eta +1}={n+1 \over n -2} \nonumber \\
\alpha_2 &=&  6 { \eta -1  \over (2 \eta + 1)^2  } {\ln ( {\cal M}_*)\over \sigma^2}=  3 { n - 5 \over (n-2)^2  }{ \ln ( {\cal M}_*)\over \sigma^2},
\label{expo}
\end{eqnarray}
yielding $\alpha_1\simeq 2.66$, $\alpha_2\simeq -1.11\,(\ln  {\cal M}_*)/ \sigma^2$ for $n=3.8$.

\noindent For usual MW molecular cloud conditions, $\mach\lesssim 10$ and $\ms^2\sim 2$, the contribution from $\alpha_2$ is not negligible and decreases the slope to a Salpeter value, $\alpha\simeq2.35$. 
In contrast, for very dense ($\lj \ll L_c$) and very turbulent ($\mach \gg 1$) conditions, the second contribution has a weaker impact, so the slope is steeper than the canonical Salpeter one.
It is interesting to examine under
which conditions this second contribution provides only a negligible correction to the first one, so that the slope remains close to $\alpha_1$. Clearly this occurs for very
large Mach numbers but also implies a condition on $\ms$, thus on the density. Using
the definitions of $\machs$ and $\sigma$, a value $|\alpha_2| \lesssim 1\%\,\alpha_1$, for instance, yields as a rough condition
 (under the condition $b\mach \gg 1$):

\begin{eqnarray}
{\bar n}  \gtrsim (4.0 \times 10^3)\,({T \over 10\,{\rm K}})\,({L_c \over 10\,{\rm pc}})^{-2}\,({\mach \over 10})^{4.8}\,\,{\rm cm}^{-3}
\label{nlim}
\end{eqnarray}
For $T=40$ K, $L_c=100$ pc, $\mach=50$, this corresponds to $\nbar \gtrsim 4 \times 10^5\cc$, 
 for $T=60$ K, $L_c=100$ pc, $\mach=60$, to $\nbar \gtrsim 10^6\cc$, and for $T=80$ K, $L_c=100$ pc, $\mach=70$, to $\nbar \gtrsim 3.5 \times 10^6\cc$.
Therefore, according to the present theory, in highly turbulent {\it and} very dense regions, 
the high-mass slope of the IMF is expected to be steeper than the Salpeter value, 
reaching up to a value $\alpha = \alpha_1 \sim 2.7$.
Not only such a steepening of the IMF in massive ETGs has been suggested by various studies (see \S\ref{intro}) but this maximum value for the high-mass slope
is in remarkable agreement with values inferred from recent spectroscopic observations \citep{Spiniello2012} and 
dynamical determinations \citep{Cappellari2012,Cappellari2013,Barnabe2013}, which exclude slopes steeper than $\alpha\approx 2.8$ for ETGs with velocity dispersion in the range 200-335$\kms$.

In contrast, in low-density and weakly turbulent environments, the high-mass tail of the IMF can be shallower than the Salpeter value (see HC09 Fig. 8). Such environments, however, barely form stars.
Indeed, there have been claims in the literature that some dwarf galaxies, characterized by very low velocity dispersions ($<10 \kms$) and SFRs 4 to 5 orders of magnitude smaller than in the MW, may have a {\it shallower} than Salpeter IMF slope \citep[e.g.][]{Geha2013}. Although consistent with the general picture described by the HC theory, as just mentioned, these results must be taken with caution. The \citet{Geha2013} analysis relies on 
the study of a very narrow stellar mass range, namely 0.5 to 0.8 $\msol$ and drawing conclusions on the general behavior of the IMF, in particular based on a single powerlaw fit, is extremely uncertain.
As clearly seen in the rightmost panel of figure 5 of that paper, narrow observed stellar mass ranges tend to predict shallower mass slopes, which might suggest a bias caused by limited statistics.
This analysis must be confirmed over a more significant mass range before robust conclusions can be drawn.

The characteristic mass of the IMF, i.e. the most probable mass for collapse, is given by the peak of the IMF ($d{\cal N}(\mtil) /d\widetilde{M} =0)$, which yields\footnote{For sake of simplicity, we assume that the peak occurs in the purely thermal regime. A finite contribution from turbulent velocity dispersion will shift the location toward slightly smaller masses.}:
 
\noindent
\begin{eqnarray}
 \widetilde{M}_{\rm peak}  =  e ^{-\sigma^2}  \Rightarrow  M_{\rm peak} = { \mj \over [1 +(b{\cal M})^2]^{a}}  \nonumber  \\ 
\simeq 0.8\,({a_J\over C_m})({\mu\over 2.33})^{-2}  ({T\over 10\,{\rm K}})^{3/2} { [{\nbar(L_c)\over 10^4\,\gcc}]^{-1/2} \over
[1+ b^2({\vrms(L_c)\over C_s})^2] }\,\msol 
    \nonumber  \\ 
 \approx_{_{(b{\cal M} \gg 1)}}  \,{\mj \over (b\mach)^{2}}, \nonumber \\
\label{mpeak}
\end{eqnarray}
where $b$ is the turbulence forcing parameter which enters eqn.(\ref{sigma0}), $a=[1-(R/L_c)^{(n^\prime-3)}]$ (see eqn.~(\ref{sigma})), 
 and $T$, $\nbar$ and $\vrms$ denote the typical temperature, mean density and large-scale velocity dispersion for a cloud of size $L_c$ (mass $M_c$), as given by eqns.(\ref{L1}), (\ref{L2}) and (\ref{mcloud}). The peak of the IMF thus also occurs at smaller masses the larger the typical Mach value of the cloud. It is interesting to examine the dependence of the peak of the IMF, $M_{\rm peak}$, upon the cloud's mass, $M_c$, according to eqn.(\ref{mpeak}). Combining this equation and equations (\ref{L1}), (\ref{L2}), (\ref{mjeans}) and $M_c\approx \rbar L_c^3\propto d_0L_c^{3-\eta_d}$, one gets

\begin{eqnarray}
M_{\rm peak}&\propto& T^{5/2}d_0^{-0.5}V_0^{-2}L_c^{\eta_d/2-2\eta}\propto L_c^{-0.3}-L_c^{-0.45} \nonumber \\
M_{\rm peak} &\propto& T^{5/2}d_0^{-0.3}V_0^{-2}M_c^{ {\eta_d/2-2\eta\over 3-\eta_d} } \propto M_c^{-0.15}-M_c^{-0.2},\nonumber \\
\label{scaling}
\end{eqnarray}
where we have used $\eta=0.4$ and where the two exponents for $L_c$ and $M_c$ correspond to $\eta_d=1$ and 0.7, respectively. As noted in HC08 (their \S7.1.4), the weak dependence of the peak mass of the IMF upon the cloud's mass (a factor 100 in
mass yields a factor $\sim 2$ in $M_{\rm peak}$) certainly partly explains the observed universality of the IMF for similar density, temperature and velocity dispersion conditions, i.e. similar $T$, $d_0$ and $V_0$.

Therefore, a major consequence of the concept of turbulent Jeans mass is that the characteristic scale/mass for fragmentation 
in a turbulent medium 
does not simply depend on the gas mean density and temperature, as in the standard Jeans mass concept of purely gravitational fragmentation, 
but depends also strongly on the Mach number. This again illustrates the respective roles of compressive turbulence motions, 
which generate the {\it initial field} of density fluctuations in the cloud, and
gravity, which introduces a characteristic (Jeans) scale for gravitational instability.
Physically speaking, comparing eqns.~(\ref{mjeans}) and (\ref{mpeak}), one can understand this result as the fact that the proper typical Jeans scale for fragmentation in 	a turbulent medium is no longer 
the one evaluated at the cloud's mean density but the one at the cloud's density compressed to higher values at all scales by the cascade of shock dominated turbulence, 
$\bar{\rho}\times[1 +(b{\cal M})^2]^{a}$. This differs drastically from star formation theories invoking only gravitational fragmentation \citep[e.g.][]{Larson2005}, characterized only by the  thermal Jeans mass. In this latter case, $M_{peak}\approx M_J\approx T^{3/2}/\rbar^{1/2}$, leading to a strong dependence of fragmentation upon gas temperature. In that case, one would expect systematically
bottom-light IMF's in warm $(T>10$ K)  environments. In the present theory, however, this temperature dependence is largely counterbalanced by the Mach dependence. 
As mentioned above, the reason for the "universality" of the IMF under Milky Way like conditions is the modest dispersion around the normalization values at 1 kpc, set up by the ones at galactic scale, which are very similar for local
galaxies, corresponding to a typical accretion rate $\sim 2\,\acc$ from the intergalactic medium \citep{Klessen2010}.

As illustrated by the above relations, the HC theory of the IMF thus naturally predicts that the IMF of {\it very dense and highly turbulent} environments should have:

- a characteristic (peak) mass shifted toward smaller masses compared with the standard MW Chabrier IMF, 

- a high-mass slope which can be steeper than the Salpeter value.
As mentioned above, this larger fraction of low-mass cores in such environments is a direct consequence of (i) the enhanced gas compression by highly turbulent motions, 
and (ii) the shorter free-fall times
for the collapsing overdense regions, increasing the relative fraction of small cores over massive ones, a process accounted for in the HC time-dependent formalism.


\section{Initial mass functions}
\label{IMF}


\subsection{Variations of the initial mass function}
\label{IMFvar}

\begin{figure*} 
\center
\includegraphics[width=12cm]{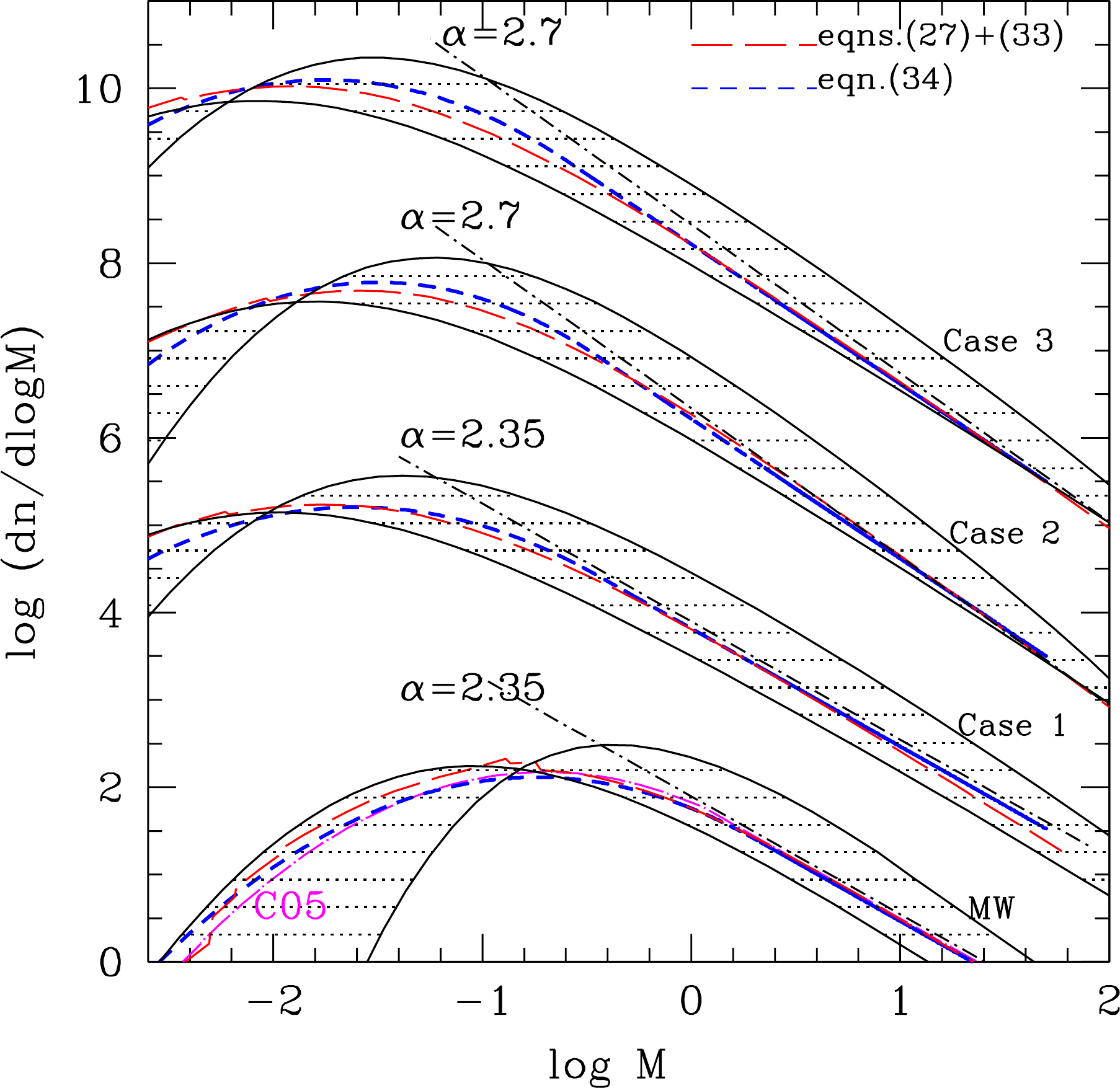}
\caption{Initial mass function according to eqn.~(\ref{grav_tot2}) for conditions corresponding to the MW case and Cases 1, 2, 3, respectively from bottom to top (masses in $\msol$).
In each case, the two solid curves correspond to single clumps of size $L=1$ pc (rightmost) and 50 pc (leftmost), respectively, for the MW, and $L=1$ and 100 pc for the other cases,
while the long-dashed (red) curves portray the integrated IMF for a clump mass distribution (eqn.~(\ref{IMFglobal})) 
and the short-dashed (blue) curves correspond to the parametrized IMF's (eqn.~(\ref{IMFequ})).
 The dot-dashed lines correspond to the \citet{Chabrier2005} object IMF (bottom, magenta, labeled C05), the Salpeter (1955) IMF ($\alpha=2.35$) and to power-law mass functions $dn/d{M}\propto M^{-2.7}$, 
the expected steepest slope according to  eqn.~(\ref{expo}). For sake of clarity, each group of curves for a given case has been shifted upward.}
\label{plotIMF}
\end{figure*}

Table \ref{Tablecas} displays 4 typical star-forming cloud conditions characterized by different gas temperature, density and velocity normalization values, $d_0$, $V_0$, 
as inferred in \S\ref{ETG}, within the expected range of cloud sizes. 
The characteristic virial parameter, 
$\alpha_{vir}=2E_K/E_G=(5/\pi) V_{{\rm rms}}^2 / (G {\bar \rho} L_c^2)$,
measures the ratio of turbulence over gravitational energy within the clump.
The case labeled "MW" is representative of typical MW conditions, with cloud sizes $L_c\simeq 1$ to 50 pc.
Cases 1, 2 and 3 should be representative of the conditions encountered in the high-redshift progenitors of massive ETGs and starburst environments, as discussed in \S3, with enhanced gas temperatures, mean densities and dispersion velocities, for cloud sizes $L_c\simeq 1$ to 100 pc. 
Figure \ref{plotIMF} portrays the corresponding IMFs, calculated with eqn~(\ref{grav_tot2}) with $b=0.5$, the value inferred in simulations for high Mach numbers \citep{Federrath2013,Kritsuk2013}. 
For each case, the two solid lines display the IMFs corresponding to the 
aforementioned bracketing cloud sizes. 
As mentioned earlier, in ULIRG-type galaxy mergers, the clouds 
form a nearly continuous medium rather than an ensemble of individual entities (Downes \& Solomon 1998). For such a case of spatially close dense clumps, there is no need to sum up over a clump population, as the HC theory naturally takes into
account the clumpy structure of the gas, and the IMF for a given typical cloud size should be representative of the galaxy-wide IMF for a galaxy of similar typical scale height.
In less dense environments, where clumps can be spatially well separated by diffuse gas, however, one must sum up over the clump population. This is illustrated by the long-dashed lines, 
which correspond to a global IMF integrated over a mass spectrum ${\cal N}_c=dn/dM_c$ of molecular clumps, as given by eq.(16) of HC08.
We recall that the mass spectrum ${\cal N}_c$ obtained in the HC theory accurately recovers the one observed for CO clumps or infrared dark clouds in the MW,
$dN_c/dM_c \propto M_c^{-1.7}$ \citep[e.g.][]{Heithausen1998,Kramer1998,Peretto2010} (see HC08).
This yields for the clump-integrated IMF:

\begin{eqnarray}
{\cal N}_{tot}  = \int_{M_{c}^{inf}}^{M_{c}^{sup}}\,{\cal N}(M_c)\,V_c\,{\cal N}_c\,dM_c, 
\label{IMFglobal}
\end{eqnarray}
where ${\cal N}(M_c)$
is the mass spectrum of self gravitating cores (eqn~(\ref{grav_tot2}))
for a clump of
mass $M_c\simeq (\pi/6)\rbar L_c^3$
and  volume $V_c$. The limits
$M_{c}^{inf}$ and $M_{c}^{sup}$ correspond to the minimum and maximum mass for these clumps according to the Larson relations (eqns.~(\ref{L1})) for
the clump sizes and density normalizations given in Table 1.

\setlength{\unitlength}{1cm}

\setlength{\unitlength}{1cm}
\begin{figure*} 
\center
\includegraphics[width=8cm]{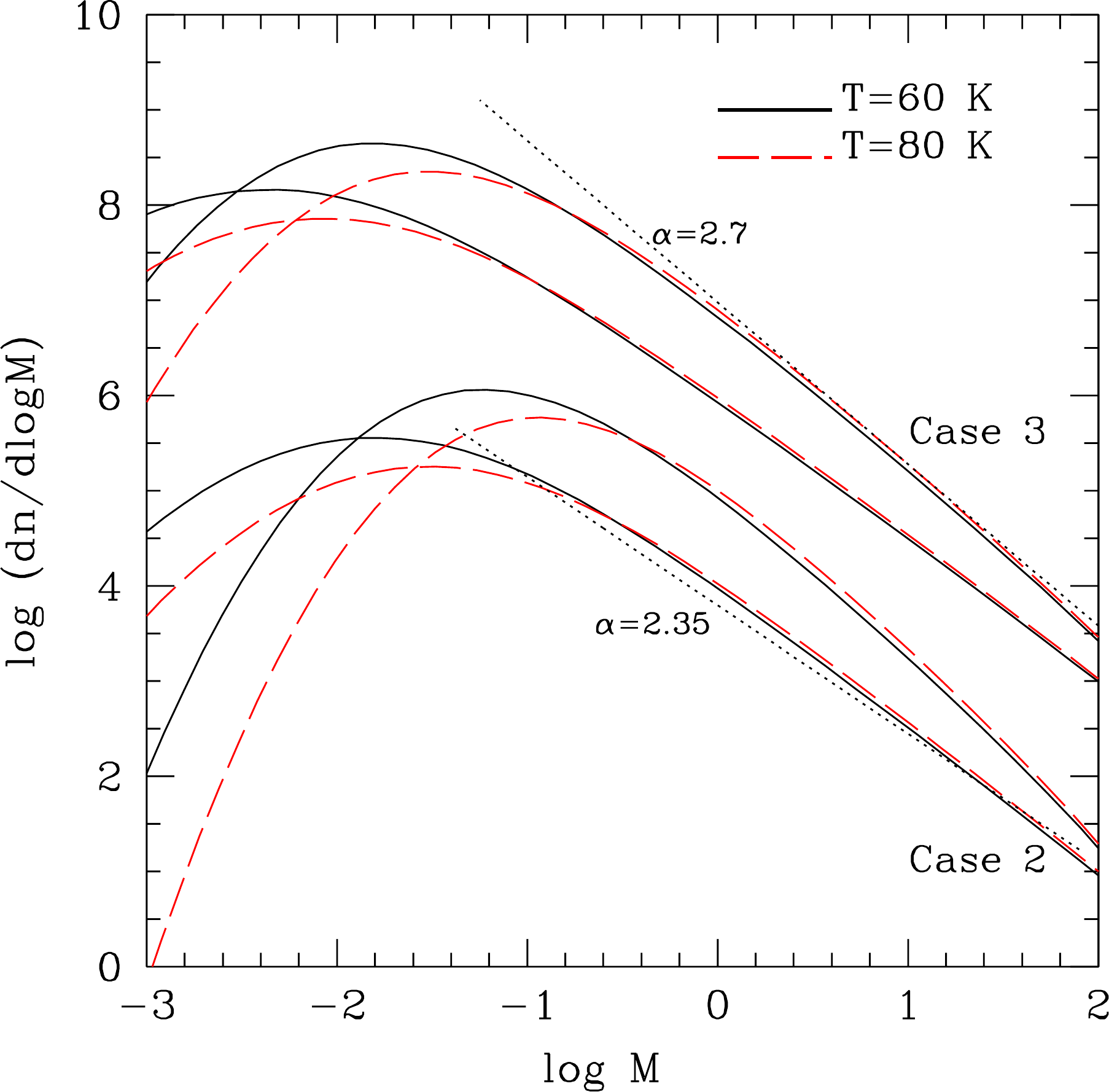}
\caption{Initial mass function as in Figure \ref{plotIMF} for Cases 2 (bottom) and 3 (top) for 2 cloud sizes, namely $L_c=1$ pc (right curves) and $L_c=$ 100 pc (left curves), for the same density conditions 
as given in Table 1 but different temperatures, namely $T=80$ K  
and $T=60$ K. }
\label{plotIMF_T}
\end{figure*}
 
For sake of simplicity and in order to focus on the very issue explored in the present paper, namely variations of the IMF under non MW-like conditions, 
the present calculations have been conducted for the case of an isothermal gas, for the respective temperatures given in Table I. 
As shown in HC09, taking into account the thermodynamics of the gas will extend the IMF into the low-mass domain for an adiabatic index $\gamma<1$.
 Given our lack of a precise
 knowledge of the dominant heating and cooling 
mechanisms in extreme starburst environments and given the expected large variety of conditions in such regions, it seems reasonable to stick for now to the simplest assumption. For the same reason, we do not explore the modification of the IMF due to binaries so the IMFs correspond to unresolved stellar systems. 
 The IMF of unresolved systems and the one for individual objects are very similar in the stellar regime \citep{Chabrier2005}. Notable 
differences start to emerge in the brown dwarf regime, unobservable in ETGs. 
Corrections due to binaries, although affecting to some level the IMFs displayed in Fig. \ref{plotIMF}, should thus
remain modest in the present context, certainly well within the uncertainties pertaining to the exact density, turbulence and temperature conditions for extreme environments.
At last, the PDF of turbulence in the present calculations is supposed to obey a lognormal form, with the same value $n^\prime=n$ for the respective
 indices of the log-density and velocity power spectra. Although, as mentioned earlier, these behaviours are verified in numerical simulations for moderate Mach values 
($\mach \lesssim 10$), they certainly become more dubious for higher values \citep[e.g.][]{Kowal2007,FederrathKlessen2013}. 
Again, the impact of the departure from these approximations on the results is likely to remain well within the range of uncertainties and expected variations characteristic of extreme
star-forming conditions. We also recall that the HC theory does not depend on any specific PDF, although a lognormal form allows a fully analytical derivation.

The \citet{Chabrier2005} IMF is shown for comparison in the figure, as well as the Salpeter IMF 
($dn/d\ln M \propto M^{-(\alpha-1)}$, with $\alpha=2.35$) and a power law IMF with $\alpha=2.7$.
Even though the general behavior of the IMF remains the same for all conditions, with the combination of a power law at large masses and a lognormal form at smaller masses, 
large Mach values and high densities shift the peak towards smaller masses compared with the MW case, with the power law part extending eventually 
down to the hydrogen burning limit, $\sim 0.1\,\msol$, in spite of the significantly larger gas temperatures. 
This reflects the general analysis carried out in the previous section.
As mentioned earlier, in the standard Jeans gravitational theory for fragmentation, such high temperatures would predict the opposite behavior, as
the characteristic mass for fragmentation, namely the mean {\it thermal} Jeans mass, would be shifted towards higher masses.
As seen in the figure, for Case 2 and more notably Case 3, which approach or fullfil the condition (\ref{nlim}), 
the IMF can get steeper than the traditional Salpeter value over a significant mass range, reaching
a slope value $\alpha\sim 2.7$, as expected from our previous analysis.

As mentioned in  \S\ref{temp}, the temperature of the gas in the progenitors of ETGs and in starburst environments is rather uncertain so it is interesting to
examine the impact of such an uncertainty upon the resulting IMF. This is illustrated in Figure \ref{plotIMF_T} where we compare the IMFs obtained with Cases 2 and 3 under the same density conditions but with different temperatures, namely $T=80$ K for Case 2 and  $T=60$ K for Case 3. 
As expected, a warmer medium implies a larger thermal Jeans mass and a lower Mach number (larger speed of sound), slightly shifting the characteristic mass of the IMF towards larger masses, as expected from  eqn.~(\ref{mpeak}).
The shift, however, remains relatively modest compared with the one induced by the Mach number dependence.

{\it The case of globular clusters}. An interesting issue can be raised at this stage, concerning globular clusters (GC). Indeed, although GCs are thought to have formed also under extreme conditions (high densities, large Mach numbers), their stellar mass function is not bottom-heavy \citep[see e.g.][Fig. 9]{Chabrier2003}.
The formation of GCs, however, is still an unsetlled issue \citep[see e.g.][]{Renzini2008}. There is observational both photometric and spectroscopic evidence that GC formation occurred through a series of multiple stellar generations. Whether this happened in a short series of successive bursts or in a more continuous process is unclear, but the implication is star formation episodes in an environment already inhabited by previous stellar populations. 
This mode of star formation differs from the case of GMCs devoid of pre-existing stars, as examined in the present context. 
Furthermore, GCs have experienced dynamical evolution and tidal interactions, leading to a mass segregation of stars with time and space and yielding notably a deficiency of low-mass stars \citep{Baumgardt2003,Paust2010}. Recovering the exact IMF from the observed present day mass function in GCs
requires dedicated dynamical evolution calculations, which will be highly valuable to explore.



\subsection{Large brown dwarf populations ?}
\label{BD}

Looking at the bottom-heavy IMFs displayed in Fig. \ref{plotIMF}, one may wonder whether the IMF extends to very low masses, suggesting the presence of a 
significantly larger
brown dwarf population in these galaxies than in the MW, which has a star-to-brown dwarf ratio $N_\star/N_{BD}\sim 4$ \citep[see][for a recent review]{Chabrier2014}. 
The bottom of the mass function, however, might be truncated at higher masses than under MW-like conditions,
precluding such a large brown dwarf population. Indeed, the IMF extends by definition down to the minimum mass for fragmentation, $M_{min}$, 
defined as
the density at which the balance between compressional heating and radiative cooling in a collapsing cloud, which ensures isothermality, breaks down. 
At this point, heating by compression of the gas is so effective against cooling as it stops the collapse, leading to the formation of a central adiabatic core.
As shown by Masunaga \& Inutsuka (1999), however, for a given mean opacity, the dependence of $M_{min}$ upon the cloud temperature changes not only quantitatively but also {\it qualitatively} respectively below and above some 
temperature. For an opacity $\kappa\sim 0.01$ cm$^2$g$^{-1}$, which corresponds to dust continuum opacity (the usually dominant coolant in the ISM), which is independent of the velocity structure of the cloud (in contrast to line cooling),
 this change in behaviour occurs around $T\sim 30$ K (Fig. 3 of Masunaga \& Inutsuka 1999). While $M_{min}$ decreases with increasing $T$ below this value, it increases with $T$ above it.
The physical reason for this behaviour is that, if the medium is warm enough, the radiative cooling rate is too large to be balanced by compressional heating and thus the gas in the collapsing core becomes optically thick before isothermality breaks down, in contrast to what occurs in a much cooler medium. 
At this stage cooling becomes ensured by radiative {\it diffusion}, which drastically reduces the radiative cooling rate ($\Lambda_{diff}\propto t_{diff}^{-1}\propto {\bar \tau}^{-2}$, where ${\bar \tau}$ is the gas optical depth). This increases by the same amount the rate at which the gravitational energy of the collapsing core is transported outward before being radiated away, then halting the collapse at larger $M_{min}$.
According to Masunaga \& Inutsuka, in this regime $M_{min}$ depends on temperature as $M_{min}\propto (T/10 \,{\rm K})^{(5+2\zeta)/6}$, where 
$\zeta$ determines the temperature dependence of the opacity, $\kappa \propto \kappa_0\times(T/10 {\rm K})^{\zeta}$ and $\zeta\simeq 2$ in this temperature regime (Bell \& Lin 1994). This yields about a factor $\sim \times 10$ increase of $M_{min}$ for $T\sim 40$ K, and
a factor $\sim \times 20$ for $T\sim 70$ K, yielding in this latter case $M_{min}\approx 0.1\msol$, about the hydrogen burning limit.
Moreover, above the same temperature limit, $M_{min}$ is found to increase with opacity as $M_{min}\propto \kappa^{1/3}$. As the metallicity observed in ETGs today is slightly oversolar, with $[\alpha/Fe]\simeq 0.2$-0.5 \citep{Thomas2005}, the opacity in these environments is expected to be larger than for MW GMCs, increasing further
$M_{min}$. Finally, under such hot and dense environments, compressional heating of the gas becomes increasingly effective ($\Gamma_g\propto C_S^2{\sqrt \rho})$ to heat up the cloud sufficiently against cooling. Therefore, given the expected larger gas
temperatures, densities and opacities in the progenitors of ETGs or in starburst environments, we expect the minimum mass for fragmentation to be substantially larger than under standard MW-like conditions, truncating the IMF at a mass limit closer to the hydrogen-burning limit.

Interestingly enough, recent observations combining gravitational lensing, stellar dynamics and spectroscopic analysis of two massive early-type lens galaxies,
yielding constraints on the total and stellar masses, respectively, suggest 
a Salpeter-like IMF over the entire stellar regime but with a low-mass limit $\sim 0.12 \,\msol$ \citep{Barnabe2013}. If confirmed, these observations 
bring support to the above analysis.


\subsection{Parametrization of the initial mass function}

For practical purposes, the IMFs given by eqn~(\ref{grav_tot2}) for the various cases displayed on the figure can be parametrized under a Chabrier-like form,
i.e. a combination of a lognormal and a powerlaw respectively below and above a typical mass $m_0$, the proper form of the IMF according to
the present gravoturbulent picture of star formation. However, in order to insure continuity of the derivative of the IMF, we
slightly modify the original form according to the suggestion of \citet{vanDokkum2008}, as:

{\small
 \begin{equation}
\xi(m)={dn\over d\log m}=
\left\{
\begin{array}{ll}
A_l\,m_0^{-x}\,\exp[-({\log m-\log m_c)^2\over 2\,\sigma^2}], \hskip0.5cm m\le m_0\\
A_h\,m^{-x}\hskip3.6cm m\ge m_0,
\end{array}
\right.
\label{IMFequ}
\end{equation}
}\\
with $m_0=n_cm_c$. Continuity of the function and its derivative implies the condition $\sigma^2=\log n_c/(x\ln 10)$ for the variance and $A_l/A_h=n_c^{x/2}$ for the normalizing coefficients. 
Table \ref{TableIMF} gives the values of $m_0$, $n_c$, $\sigma$ and $x$ for the different cases examined in the present study, as well as the peak mass for the corresponding mass spectrum, $dN/dM$.
These parametrized IMFs are illustrated by the short-dashed lines in Fig. \ref{plotIMF}.
A seen in the figure, the parametrized IMF labeled "MW" in Table 2 is very similar to the standard \citet{Chabrier2005} IMF for resolved objects, characteristic of the MW environment. 
The number and mass integrals can easily be calculated analytically:

{\tiny
 \begin{eqnarray}
{\mathbb N} &=& \int_{m_{inf}}^{m_{sup}}\xi(m)\,d\log m \nonumber \\
& = &{ {\sqrt{2\pi}} \over 2}\, \sigma \,(A_l\,m_0^{-x})\, \times \Bigl\{{\rm erf} \Big( {\log m_{0}-\log m_c)\over {\sqrt 2}\,\sigma} \Big) - 
{\rm erf}\Big( {\log m_{inf}-\log m_c) \over {\sqrt 2}\,\sigma} \Big) \Bigr\} \nonumber \\
&-& ({A_h\over x\cdot \ln 10})\,\big (m_{sup}^{-x}-m_{0}^{-x} \big)
\label{integnum} \\ \nonumber \\
{\mathbb M} &=& \int_{m_{inf}}^{m_{sup}}m\,\xi(m)\,d\log m \nonumber \\
 &=& { {\sqrt{2\pi}} \over 2}\, \sigma \,(A_l\,m_0^{-x})\,\exp(y^\prime_c+{\sigma^{\prime^2}\over 2}) \times \Bigl\{{\rm erf} (X_0) - {\rm erf}(X_{inf}) \Bigr\} \nonumber \\
&+& \Big({A_h\over (1-x)\cdot\ln 10}\Big)\,\big (m_{sup}^{1-x}-m_{0}^{1-x} \big),
\label{integmas}
\end{eqnarray}
}
where $y^\prime_i=(\log m_i)\times \ln 10$, $\sigma^\prime=\sigma\times \ln 10$ and $X_i=[y_i^\prime-(y^\prime_c+\sigma^{\prime^2})]/({\sqrt 2}\,\sigma^\prime)$.
These parametrizations will be useful in galaxy evolution calculations
 to explore the consequences of varying IMFs under non standard star-forming conditions such as the ones explored in the present study.


\section{Star formation rates}
\label{SFR}

The star formation rates obtained with these different IMFs are calculated as in \citet{Hennebelle2011a,HennebelleChabrier2013}: 

\begin{eqnarray}
SFR_{ff} = \epsilon \, \int _0 ^{M_{cut}}  { M {\cal N} (M) dM \over \bar{\rho}},
\label{SFRff}
\end{eqnarray} 
where ${\cal N}(M)$ is the (time-dependent) IMF of prestellar cores given by  eqn~(\ref{grav_tot2}) and $M_{cut}$ is the largest unstable mass in the cloud, typically a fraction of this latter (see HC13).
The parameter $\epsilon =M_\star/M_{core}$ is the efficiency with which the mass within
the collapsing prestellar core is converted into stars, i.e. the fraction of the prestellar core infalling gas effectively accreted by the nascent star.
The other fraction is expelled
by jets and outflows during the collapse (see \S\ref{HC}). Then, $\epsilon$ represents the {\it local} core-to-star formation efficiency. 
Calculations \citep[e.g.][]{Matzner2000,Ciardi2010} as well as observations \citep[e.g.][]{Andre2010} 
suggest a value $\epsilon \simeq 0.3-0.7$, yielding a factor $(\epsilon/\phi_t)\approx 0.1$-0.3. 
Note that $SFR_{ff}$ is a dimensionless quantity, namely the star formation rate per cloud mean freefall time $\tau_{ff}^0=(3\pi/32 G \rbar)^{1/2}$,
i.e. the fraction
of cloud mass converted into stars per cloud mean freefall time: $SFR_{ff}=({\dot M}_\star/M_c)\tau_{ff}^0$ \citep{Krumholz2005,Hennebelle2011a,HennebelleChabrier2013,Federrath2013}.
 For star-forming
galaxies, supposed to be marginally Toomre stable, $Q\approx 1$, this time-scale is about the disk orbital period \citep{Krumholz2005,Krumholz2012}. 
The {\it global} star formation rate at the global {\it cloud} scale, however,
must also include the {\it global efficiency}, $SFE$, i.e. the typical fraction of star-forming (essentially molecular) gas effectively forming stars within clouds (i.e. within the galaxy, 
assuming all galactic molecular gas resides in clouds).
Observations \citep{Evans2009,Lada2010} as
well as simulations 
\citep{FederrathKlessen2013} suggest $SFE$ $\approx 1\%$-6\% within GMCs, and a similar value for the galaxy-averaged star formation efficiency \citep{Kennicutt1998,Swinbank2011}. This yields a {\it global} 
star formation rate 

\begin{eqnarray}
 SFR=SFE\times SFR_{ff},
\label{SFRSFE} 
\end{eqnarray}
 and thus a total volume and projected SFR densities, respectively:

\begin{eqnarray}
{\dot \rho}_\star&=&{\rho_\star \over \tau_{ff}^0}=SFR\,\times({\rho_g \over \tau_{ff}^0}),\label{SFRvol} \\
{\dot \Sigma}_\star&=&{\Sigma_\star \over \tau_{ff}^0}=SFR\,\times({\Sigma_g \over \tau_{ff}^0}),\label{SFRpro}
\end{eqnarray}
where ${\Sigma}_g$ is here the gas {\it projected} density evaluated for the area under consideration, be it a GMC of a galaxy.
For a roughly homogeneous spherical cloud of size $L_c$, $\Sigma_g\approx {\rbar}\,L_c$, while for a galaxy of typical disk scale height $h$, 
$\Sigma_g\approx 2\,{\rbar}\,h$.
Clearly, the determinations of both the average free-fall time and gas density retain significant ambiguities \citep[see e.g.][]{Krumholz2012,Evans2014}. 
In particular, they both involve some scale over which they are averaged, assuming the gas distribution is uniform over this scale.
As previously, we assume for simplicity that all the galactic molecular gas resides in the GMCs, whose maximum size 
is the typical injection scale of turbulence, i.e. the galactic scale height (\S 2.2), so that $L_c\sim h$ \citep[see e.g.][]{Krumholz2005}. 
Note that in case these two scales differ appreciably,
with $L_c\ll h$, eqn.(\ref{SFRpro}) implies a dependence on the scale height as ${\dot \Sigma}_\star\propto (L_c/h)^{1/2}$. 

\begin{figure*} 
\center
\includegraphics[width=12cm]{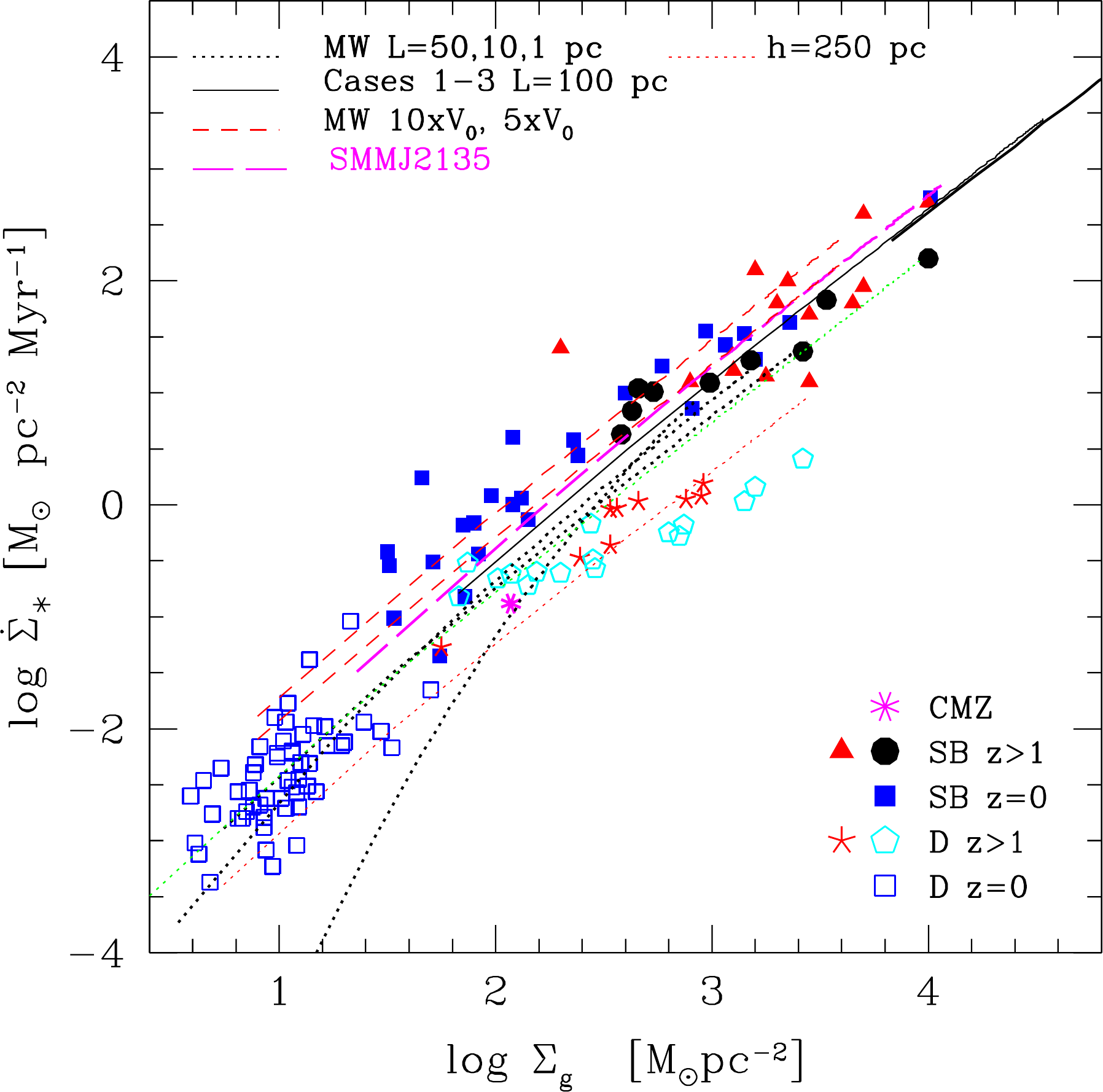}
\caption{{\footnotesize Global star formation rates as a function of gas surface density for various conditions as obtained by eqn.(\ref{SFRff}),
assuming a global star formation efficiency $SFE=1\%$. Thick (black) dotted lines: MW-like conditions in Table 1 for cloud sizes $L_c=$1, 10 and 50 pc, 
from bottom to top. Thin (red) dotted line: MW-like condition for a cloud size $L_c=50$ pc but a scale height $h=250$ pc (see text). Superposed 
(barely distinguishable) black solid lines: density and velocity normalization conditions ($d_0$, $V_0$) corresponding to Cases 1 to 3 over a large density range
 for cloud size $L_c=100$ pc. The corresponding Mach values, $\mach$, are given in Table 1.
Long-dash line: conditions characteristic of SMMJ2135. 
Short-dash lines: MW-like {\it density} ($d_0$) conditions but for normalization of the
turbulent velocity amplitude at 1 pc respectively 5 times ($5\times V_0$) (lower line) and 10 times ($10 \times V_0$) (upper line) the ones of the MW. 
For all these calculations, we have adopted the average parameter value $\epsilon/\phi_t=0.2$.
{\it Empty symbols} correspond to disk galaxies: \citet{Kennicutt1998}, blue squares; \citet{Daddi2010}, red stars; \citet{Tacconi2010}, cyan hexagons. 
Red asterisk: CMZ \citep{Yusef-Zadeh2009}. {\it Solid symbols} correspond to low-$z$ and high-$z$
starburst galaxies: Kennicutt (1998), blue squares; \citet{Bouche2007}, red triangles; \citet{Genzel2010}, solid black circles.} 
}
\label{plotSFR}
\end{figure*}

The traditional way to look at SFRs is to examine the relationship between ${\dot \Sigma}_\star$ and ${\Sigma}_g$, as illustrated by the well-known 
Kennicutt-Schmidt relation ${\dot \Sigma}_\star \propto {\Sigma}_g^{1.4}$. This is
 portrayed in Figure \ref{plotSFR}, as obtained from eqns.(\ref{SFRff}), (\ref{SFRSFE}) and (\ref{SFRpro})
for a global star formation efficiency $SFE=1\%$, for different cloud sizes (see figure caption),
 for the cloud characteristic conditions explored in the previous section\footnote{As mentioned earlier, strickly speaking, the integral in eqn.(\ref{SFRff}) in the Hennebelle-Chabrier theory involves 2 terms. 
The second term provides a truncation of the IMF 
at large scales, leading to slightly lower SFR values than the present ones. For sake of simplicity, however, this term has been dropped in the present calculations 
(see HC11 and HC13 for details).}; we have used the average value $\epsilon/\phi_t=0.2$.
Observational results are displayed for comparison.
Empty symbols correspond to SFR determinations in {\it disk} local and high-$z$ star-forming galactic regions while solid symbols portray various data observed in {\it starburst} regions and high-z galaxies 
(see figure caption). 
As mentioned above, uncertainties both in ${\dot \Sigma}_\star$ and ${\Sigma}_g$ determinations
can easily translate into at least an order of magnitude or so uncertainty on these values. 
The dotted lines correspond to the theoretical calculations for MW conditions in Table \ref{TableIMF}, for $SFE=1\%$. As shown in HC11 and HC13, the HC theory adequately reproduces the observed SFRs in MW molecular clouds 
(see e.g. Fig. 7 of HC13).

Several conclusions can be drawn from the figure. 
First, as shown in HC11 and HC13, star formation still proceeds, although at small rates, in low-density 
environments, but mostly in large enough ($L_c\gtrsim 10$ pc) clouds. Indeed, only in such clouds is turbulence strong enough (according to 
Larson relations, see Table 1) to induce dense enough gravitationally unstable density fluctuations (see HC08). 
So there is no real ``break`` or ``threshold'' in star formation but rather a continuous ``bending'' of the ${\dot \Sigma}_\star$ vs ${\Sigma}_g$ relation which seems to be
adequately reproduced by the theory. 
At higher density, such a scale dependence becomes weaker since the free-fall
timescale of most density fluctuations becomes short enough for these regions to collapse before getting a chance to be wiped out by turbulent motions \citep[see][]{Chabrier2011}.
Second, the higher SFRs in starburst (SB) systems compared with disk (D) galaxies at same gas surface density are
 well explained by the higher level of turbulence (higher Mach value) at all scales, as illustrated by Cases 1 to 3 but also by the short-dashed lines. 
These latter display our calculations for a density normalization $d_0$ at 1 pc typical of the MW conditions but with a turbulent velocity amplitude normalization, $V_0$,
5$\times$ and 10$\times$ larger than the standard value. 
Indeed, an intense star formation activity, typical of starburst conditions, implies a high rate of supernovae explosions, 
naturally increasing the level of turbulence at cloud scales. 
The (magenta) long-dash line displays the results for conditions in star-forming clumps
 for the high-z massive galaxy SMMJ2135 \citep{Swinbank2011}. Star formation densities in such environments are inferred to be about $15\pm5$ times higher than typically found
locally (\citep{Genzel2010,Swinbank2010,Swinbank2012}, in good agreement with our theoretical results.
Third, the increasing SFRs with increasing redshift reflect not only the increasing level of turbulence but also the larger density normalization $d_0$ (more compact structures), yielding
shorter dynamical times.

These results suggest that at least part of the spread in the observed SFRs stems from variations in the general level of turbulence, illustrating the important role of this
latter in star formation.
Larger Mach values yield larger gas compression and thus higher SFRs.
Indeed, both theory (HC11, HC13) and simulations \citep{Padoan2011,Federrath2012,Federrath2013} show that turbulence overall enhances star formation efficiency. 

The calculations, however, overestimate the SFRs for high-redshift {\it disk} galaxies by up to an order of magnitude. 
Besides large uncertainties both in the theoretical parameters ($\epsilon, \phi_t$) and in the observational determinations, as mentioned above, one can examine 
possible explanations for this discrepancy. The first one would be a smaller 
fraction of molecular gas ($f_{H_2}$)
in spirals than in starbursting systems, decreasing the global star formation efficiency. Indeed, while the ISM in MW-like disks is essentially atomic, 
it is almost fully molecular in ULIRGs. A global efficiency $SFE<1\%$, however, is a rather low value.
A second explanation is thicker disks in spirals compared with starbursts, with $h> 100$ pc. yielding a larger scale height and thus a lower projected SFR value.
Scale heights in ULIRGS are indeed found to be substantially smaller than
in the MW (Downes \& Solomon 1998). As mentioned above, the SFR density depends on the scale height as ${\dot \Sigma}_\star\propto h^{-1/2}$.
Therefore, using e.g. a value $h=250$ pc would shift the MW dotted line obtained for $L_c=50$ pc by a factor $\log [(L_c/2h)^{1/2}]=-0.5$.
This is llustrated by the thin red dotted line in the figure. In that case, the theoretical relation passes through the data at high density but lies at the very lower edge of the ones at low 
density. Therefore, the most apparent conclusion is that there seems to be a break in the slope of the $\dot {\Sigma}_\star$ vs ${\Sigma}_g$ relation for disk galaxies near a gas
surface density $\Sigma_g\approx 100\mpc2 $, with the slope becoming shallower above this value, in contrast to the starburst systems.

The green asterisk symbol in the figure portrays the SFR inferred for the CMZ, for the appropriate gas surface density \citep{Yusef-Zadeh2009,Kruijssen2014}.
As mentioned earlier, the velocity dispersions in this region are observed to be approximately $\sim$ 3 to $5\times$ larger than for typical GMCs 
 \citep[e.g.][]{Swinbank2011,Shetty2012,Kruijssen2013}, resembling our typical Case 1 (see Table 1).
 The SFR relation correponding to this latter case is illlustrated by the lower part of the solid black line. For a gas surface density $\Sigma_g=120 \mpc2$ and 
 a cloud size $L_c=50$ pc, the gas scale height of the CMZ \citep[Table 1 of][]{Kruijssen2014},
 the predicted SFR surface density for Case 1 with our fiducial value $SFE=1\%$ is ${\dot \Sigma}_\star \simeq  0.4\myr$, about a factor 2 to 3 larger than the observational determinations,  $0.13^{0.2}_{0.09}\myr$ 
 \citep{Yusef-Zadeh2009,Kruijssen2014}. Therefore, although the star formation rate in the CMZ is admitedly more modest than expected,
 the problem might not be as acute as previously thought.
 
\setlength{\unitlength}{1cm}

\setlength{\unitlength}{1cm}
\begin{figure*} 
\center
\includegraphics[width=12cm]{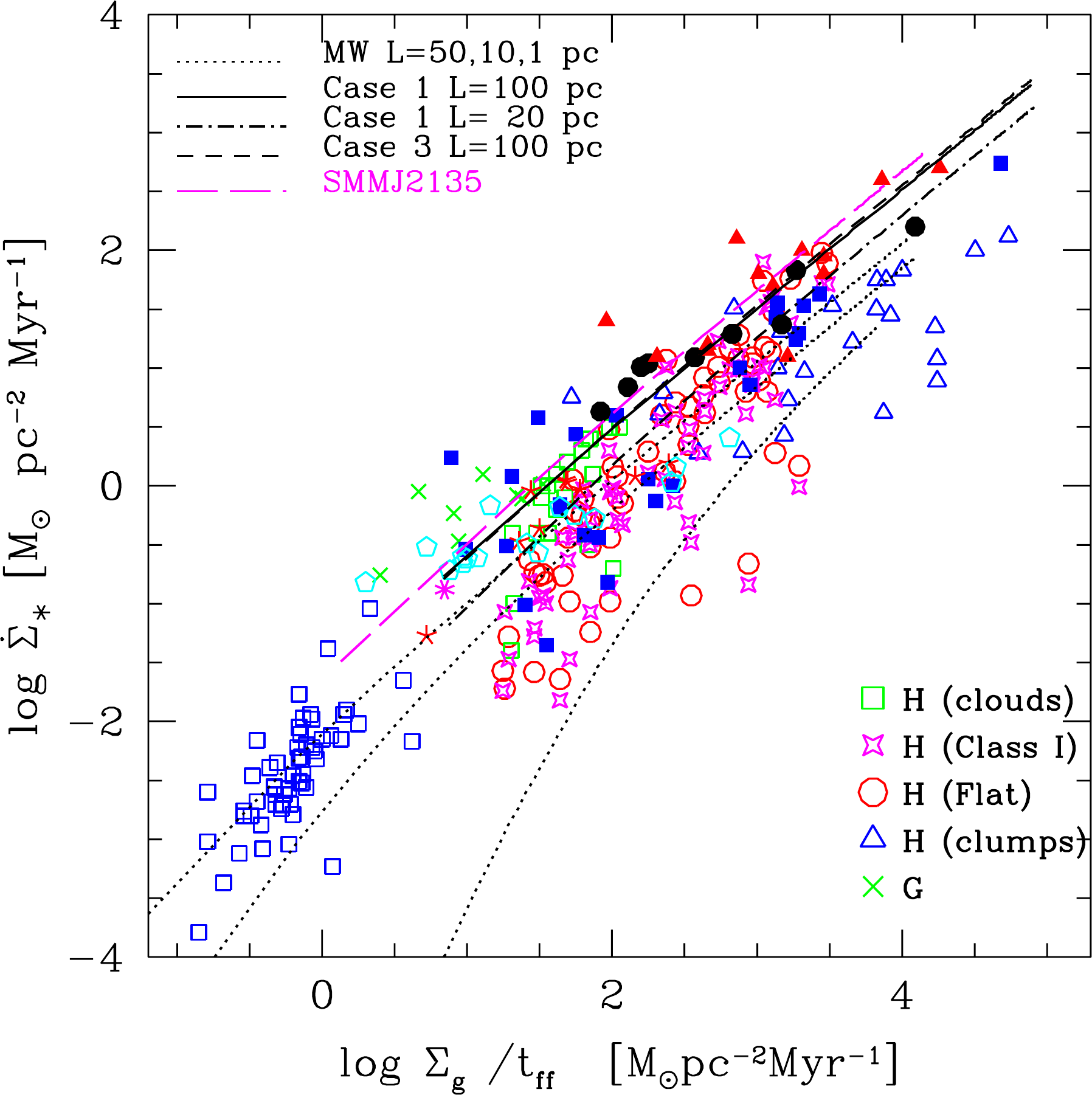}
\caption{Global star formation rates as a function of the gas surface density over dynamical time-scale (eqn.(\ref{SFRpro})) for various cloud conditions,
as displayed in the top left corner of the figure, for a global star formation efficiency $SFE=1\%$. Data symbols are the same as in Fig. \ref{plotSFR} but 
we have added data for Milky Way molecular clouds from \citet[][H]{Heiderman2010} and \citet[][G]{Gutermuth2011}.
}
\label{plotSFRtff}
\end{figure*}
 
These results clearly show the strong correlation between the star formation rate and the gas surface density, 
with an average slope consistent with the Kennicutt-Schmidt relation, ${\dot \Sigma}_\star\propto {\Sigma_g}^{1.4}$, 
as expected if $\dot{\rho_\star}\propto \rho_g/\tau_{ff}\propto \rho_g^{3/2}$. 
It has been suggested that, at high surface density, feedback, due e.g. to ionization, heating, winds etc..., takes over turbulence as the main regulator of star
formation \citep[e.g.][]{Renaud2012}. Then, the characteristic timescale for star formation is no longer the free-fall time, but a given time after which the gas becomes
available again for star formation. Since this timescale does not depend on density, this yields a linear slope $\dot{\rho_\star}\propto \rho_g$. 
Given the large spread in the observed SFRs, however, it is not possible to favor one of the two relations, except possibly, as mentioned above, for the high-$z$ disk galaxies.
Feedback at large (galactic) scale, however, certainly affects the global SFE, for instance by disrupting or photoionizing the cloud themselves, eventually shutting off
star formation above some critical formation rate.

All in all, within the aforementioned range of expected {\it global} star formation efficiencies, $SFE\approx 1\%$-6\%, and
{\it local} efficiencies, $(\epsilon/\phi_t)\approx 0.1$-0.3, i.e. a 1.2 dex vertical spread, the present calculations well bracket all SFR determinations over five orders of
magnitude in density, from local disk galaxies to high-z starburst systems.
This strongly supports 
the idea of star formation resulting from a dominant, or more exactly two dominant universal mechanisms, namely turbulence, which generates the original field
of density fluctuations, and gravity, which determines the dynamical time of these fluctuations, as described by the present general time-dependent gravo-turbulent theory. The global efficiency, $SFE$, however,
 certainly depends upon the environment conditions (or equivalently the initial conditions), notably the strength of turbulence at the injection scale and the mean density of the medium, which set up the density and velocity 
 amplitude normalizations in Larson relations at cloud scale. 


It has been suggested \citep[e.g.][KDM]{Krumholz2012} that ploting ${\dot \Sigma}_\star$ as a function of ${\Sigma_g}/\tau_{ff}$ instead of ${\Sigma_g}$ 
provides a better representation of the SFR relation than the one
displayed in Figure \ref{plotSFR}. This is illustrated in Figure \ref{plotSFRtff}, still with $SFE=1\%$ in the calculations, where the observational values, including the free-fall time (more exactly dynamical time) determinations
 are the ones determined in KDM \citep[as in][]{Federrath2013}, keeping in mind that these values retain significant uncertainties. 
We have added in this figure the data for Milky Way molecular clouds \citep{Heiderman2010,Gutermuth2011}.
The spread in ${\dot \Sigma}_\star$ is still significant. As noted above and stressed in HC11 and HC13, the
SFR not only depends upon density, but also strongly depends on the cloud's size/mass, in particular at low surface density. So there is no strickly speaking some ``universal`` star formation relationship between ${\dot \Sigma}_\star$ and ${\Sigma_g}$, with a unique exponent, but rather different relations, depending on the cloud physical properties, with significant variations leading to a large dispersion, a point also advocated by observational analysis \citep{Shetty2013,Shetty2014}.
As mentioned above, this scale dependence reflects the dominant role of large-scale turbulence as the main driver for star formation, since the Mach number increases with $L_c$ (eqn.(\ref{L2})).
There is thus a degeneracy between the level of turbulence and
the global star formation efficiency, the two quantities being interconnected.
Therefore, although the data are roughly consistent with an average value $SFR\approx 1\%$ in eqn.(\ref{SFRpro}), i.e. ${\dot \Sigma}_\star \approx 0.01\,\times({\Sigma_g / \tau_{ff}^0})$, as suggested by KDM, this global 
relationship must be taken with caution.
Indeed, such a fixed efficiency does not capture the aforementioned dependence of star formation efficiency upon turbulence and thus upon cloud's scale \citep[see e.g.][]{Evans2014}.


\section{Mass-to-light ratio}
\label{MLR}

On one hand, lensing observations
constrain the total (maximum) mass of a galaxy within its Einstein radius. This comprises the total stellar mass, including stellar remnants, the ejected gas during stellar evolution 
and the dark matter contribution. The stellar mass contribution, on the other hand, can be separated from the dark matter one with integral-field data using galaxy dynamical models \citep{Cappellari2012,Cappellari2013}. This yields 
eventually the {\it stellar} mass-to-light ratio, $\Upsilon_\star=(M/L)_\star$. 
In ETGs, these ratios have been found to be $\sim$ 2$\times$ or more larger than 
those corresponding to a Chabrier IMF, typical of MW-like galaxies \citep{Conroy2012a,Conroy2013,Cappellari2012,Cappellari2013,Tortora2013,Tortora2014,Spiniello2012,Spiniello2014}. 

\begin{figure*} 
\center
\includegraphics[width=14cm]{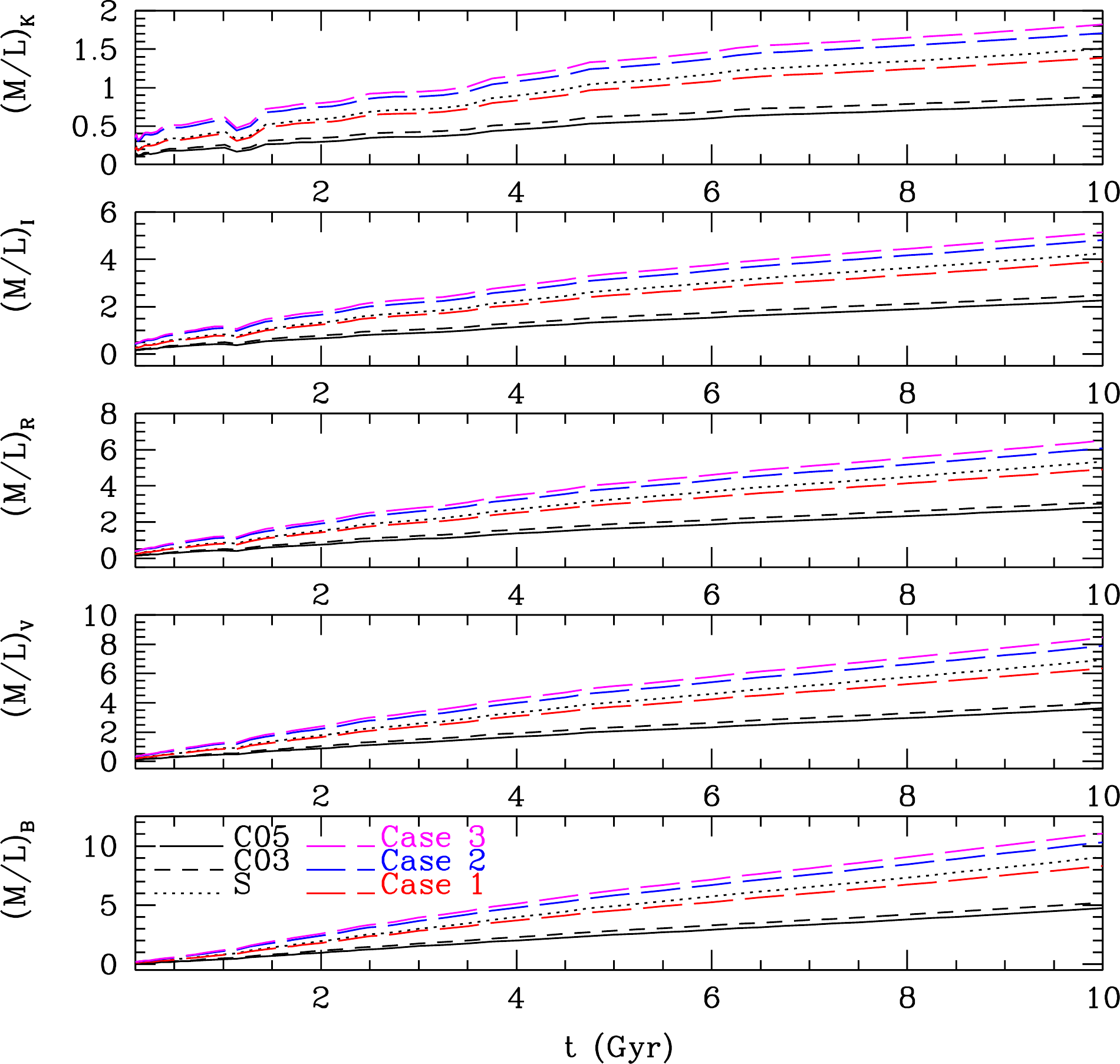}
\caption{Stellar mass-to-light ratio, $(M/L)_\star$, as a function of age in various photometric bands obtained from SSP models using
 the IMF's corresponding to: the Milky Way disk with the Chabrier 2005 IMF (solid line) or the Chabrier 2003 IMF (short-dashed line),
the Salpeter IMF (dotted line), and the IMFs corresponding to Cases 1 (red), 2 (blue) and 3 (magenta) (long-dashed lines going upward).}
\label{plotML}
\end{figure*}

As mentioned earlier, star formation in the progenitors of massive ETGs must have taken place within a relatively short timescale. This is confirmed by the fact that
the most massive galaxies, born in the highest density peaks of the primordial fluctuations, appear to be enhanced in $\alpha$-elements (\S 1 and 3).
This indicates that the duration of star formation decreases with increasing mass, having been shorter than $\sim 1$ Gyr,
 the typical timescale for the onset of Type Ia SNe, for the most massive galaxies. 
Hence, we can explore the stellar M/L ratios predicted by our formalism in these systems by using spectral evolution models for single stellar populations (SSP). 
Figure \ref{plotML} illustrates the evolution of 
$\Upsilon_\star$ as a function of age 
 in various photometric bands obtained from SSP models \citep{Bruzual2003} with the IMFs corresponding to the ones characterized in Table \ref{TableIMF} 
and with a Salpeter IMF. 
 Table \ref{TableML} gives the corresponding values
at an age of 10 Gyr, about the present age of observed ETGs. Changing the age of the stellar population in the SSP synthetic spectra 
anywhere in the range $\sim 8$ to 12 Gyr changes the $M/L$ ratio by less than 20\%. 
Note that these values represent only the {\it stellar} M/L ratios and do not include the brown dwarf contribution. 
Indeed, observationally, the stellar M/L are determined essentially from the comparison of measured equivalent widths of given spectral lines (e.g. NaI, NaD, Ti0$_2$) characteristic of the low-mass star population with those derived from SSP synthetic models computed with different IMF's, by fitting integrated spectra.
 They are thus only sensitive to
the stellar population. Moreover, as discussed in \S\ref{BD}, the mass limit for star formation in extreme environments might be truncated at a significantly larger value than in the MW, precluding 
the existence of a large brown dwarf population. 
As seen in Table \ref{TableML}, the  stellar $M/L$ ratios in various passbands are in very good agreement with the ones inferred from various observations \citep[e.g.][]{Conroy2012a,Conroy2013}.
As mentioned earlier, the present calculations do not consider corrections to the IMFs due to stellar multiplicity. In order to estimate the impact of this correction upon the M/L ratios, we have calculated these values with the \citet{Chabrier2005} IMF for unresolved {\it systems} and resolved {\it objects}, respectively. For all bands,
the difference is less than 8\%, significantly smaller than differences between the values obtained with the various IMFs. This is not surprising as fragmentation of systems into multiple objects affects essentially the low-mass part of the IMF and the effect remains modest in the {\it stellar} regime \citep[see e.g.][]{Chabrier2010}.
 Note in passing that, in the
cases of bottom heavy IMFs, there is no need for (significant) dark matter contribution, in agreement with recent observations of lens massive ETGs \citep{Barnabe2013}.

\setlength{\unitlength}{1cm}

At this stage, one should add a word of caution. As mentioned earlier, the metallicity inferred for ETGs is slightly oversolar, with $[\alpha/Fe]\simeq 0.2$-0.5. 
One cannot rule out that the SSP models used to calculate M/L ratios are not correct for such metal-rich and $\alpha$-enhanced stellar populations. One thus cannot completely exclude that the high inferred M/L values for such galaxies does not necessarily stem from a variation of the IMF but from a higher 
mass fraction under the form of stellar remnants. The integrated surface brightness of such a remnant population has been calculated by \citet{Chabrier2004} and,
although beyond present-day observational capabilities, might be detectable with the JWST. 


\section{Conclusion}
\label{conc}



In this paper, we have shown that, under extreme conditions of very dense and turbulent gas, as encountered in starbursts or in the progenitors of massive ETGs, the characteristic (peak)
mass of the IMF can extend to lower masses than under conditions typical of  most spiral galaxies, in spite of the expected higher gas temperature.
This is a direct consequence of the dominant role played by compressive turbulent motions in setting up
the very initial field of density fluctuations which determine the initial conditions for star formation and the subsequent IMF. 
Indeed, in a gravoturbulent picture of star formation, the characteristic mass for fragmentation is not the mean thermal Jeans mass, as in the classical gravitational fragmentation
scenario, but the turbulent Jeans mass, which strongly depends on the cloud's large scale Mach number. 
At small scales, large-scale turbulence cascades into local shocks that drastically increase the local gas density, which in turn triggers the collapse of dense enough regions into gravitationally bound prestellar core embryos. The consequences are a characteristic mass for the IMF in very dense and turbulent environments which shifts toward smaller masses compared with more quiescent or less dense conditions.
According to the present theory, above some density, the high-mass slope of the IMF in such environments can
 get steeper than the Salpeter value, reaching a limit $\alpha \sim 2.7$, a consequence, in particular,
 of the {time-dependence of the turbulence induced  fragmentation process, as incorporated in Hennebelle \& Chabrier's theory \citeyearpar{Hennebelle2011a,HennebelleChabrier2013}.  
The present results thus provide a theoretical foundation to observational indications that, while (less massive) spiral galaxies are consistent with a Chabrier IMF, 
more massive and dense ones, in particular massive ellipticals, require a Salpeter or even steeper IMF.

We provide simple estimates of the cloud typical properties, i.e. mean density $d_0$ and velocity dispersion $V_0$ normalizations at 1 pc, 
the relevant scale for the IMF, in terms of those at 
galactic scale. 
The characteristic mass of the IMF is thus ultimately related to the prevailing conditions in the host galaxy, temperature, surface density and velocity dispersion.
As illustrated by eqn.(\ref{scaling}), for similar ($T, d_0, V_0$) conditions, the IMF is predicted to exhibit little variations, a consequence of the similar but
opposite scaling dependences of the Jeans mass and rms velocity upon cloud's size.
The theory then naturally explains the "universality" of the IMF for {\it similar environments}, i.e. similar values of $T, d_0$ and $V_0$. 

We have parametrized IMFs representative
of various extreme star-forming conditions, to be used in stellar population calculations aimed at exploring IMF variations in various environments. 
We have also calculated the star formation rates for these systems and confirmed that denser gas and larger turbulence significantly increase the SFR. 
Star formation rates in the progenitors of massive ETGs at high-redshift should be orders of magnitude larger than at later epochs.
We have also shown that the mass-to-light ratios calculated with the IMFs representative of the examined extreme conditions are consistent with the observationally inferred values. 
We speculate that, in spite of an IMF characteristic mass extending to lower masses than for MW-like conditions, these systems do not necessarily contain a
large brown dwarf population, because for the corresponding temperatures the minimum mass for fragmentation should be significantly larger than for typical 10 K cloud conditions. Inferring the total dynamical mass, corrected for the dark matter contribution, and using one of the presently determined IMFs should enable us to confirm or not this suggestion.

The generic conclusion of the present calculations is that turbulence induced star formation is indeed a universal mechanism and that star formation in massive, elliptical galaxies proceeded similarly to within disk and spiral galaxies but that gas density was much larger {\it and} turbulence was much more vigorous due mainly to intense accretion flows or merger events. 
The present fragmentation-induced turbulence theory for star formation naturally predicts the IMF to become increasingly bottom-heavy, thus the mass-to-light ratio to increase, in such 
environments, a consequence of the combination of high turbulence and density.
We caution, however, that the evolution of the IMF does not necessarily correlate with the galaxy mass, i.e. the stellar velocity dispersion, but with its density, thus compactness, which is
correlated with its mode of formation. We thus suggest that only massive galaxies having experienced rapid
starburst episodes will have a bottom-heavy IMF, i.e. a large fraction of M-dwarfs compared with "standards". 
These starburst events are a direct
consequence of high accretion rates, due to merging events or intense gas flows, which not only increase the level of turbulence but also the density, 
thus the compactness of the galaxy. 
This might explain why the giant, {\it low density}, elliptical ESO325-G004 seems to have an IMF compatible with the MW one, in spite of its large mass and velocity dispersion ($>300 \kms$) \citep{Smith2013}.
A transition of the IMF might thus have occured between massive galaxies which formed at high redshift by merging of {\it compact} primodial structures dominated by strong gas flows and starbursts, followed by gas-poor merging, and spiral galaxies formed at lower redshift where star formation proceeded essentially internally as a continuous, quiescent process.
As the progenitors of ETG's are the dominant galaxy population at high redshift, it might be interesting to revise accordingly the star formation evolution in the universe using the various IMFs 
derived in the present study. Although it is unclear what fraction of the total star formation has taken place in such extreme environments.



\acknowledgments
The authors are grateful to C. Federrath for providing the data used in figures 3 and 4. 
This research has received funding from the European Research Council under the European
 Community's Seventh Framework Programme (FP7/2007-2013 Grant Agreements no. 247060, no. 306483 and no. 321323NEOGAL).


\newpage

\center{
\begin{deluxetable}{lcccc} 
\tabletypesize{\footnotesize}
\tablecolumns{5}
\tablewidth{0pt}
\tablecaption{Star-forming cloud conditions for a range of cloud sizes $L_c$, characterized by different temperatures and density and velocity dispersion normalizations at 1 pc, $d_0$, $V_0$. }
\tablehead{
   \colhead{}  &  \colhead{MW}   &  \colhead{Case 1}    &  \colhead{Case 2}     &\colhead{Case 3} 
   }
\startdata
T [K] &  10 & 40  &  60  &  80  \nl
 $d_0$ [$\cc$] &  $3.5\times 10^{3}$ &   $3.0\times 10^{5}$ & $1.0\times 10^{6}$  & $3.0\times 10^{6}$  \nl
 $V_0$ [$\kms$] &  $0.8$ &  $5.0$ &  $5.0$   &  $8.0$  \nl
\hline
$C_s$ [$\kms$] &  $0.19$ &  $0.38$ &  $0.46$   &  $0.53$  \nl
$L_c$ [pc] &  1 - 50 &  1 - 100 &  1 - 100   & 1 - 100  \nl
 ${\bar n}$ [$\cc$] & $3.5\times 10^{3}$ - $2.2\times10^{2}$ & $3.0\times 10^{5}$ - $1.2\times 10^{4}$ - & $1.0\times 10^{6}$ - $4.0\times10^{4}$ & $3.0\times 10^{6}$ - $1.2\times 10^{5}$  \nl
 $M_c$  [$\msol$] & $1.0\times10^2$ - $0.8\times 10^6$ & $8.9\times 10^3$ - $3.5\times 10^8$ & $3.0\times 10^4$ - $1.2 \times10^{9}$ & $9\times 10^4$ - $3.5\times 10^{9}$  \nl
 $\mj$  [$\msol$]  & 0.7-3.5 & 0.7-3.8 & 0.8-3.8 & 0.7-3.4  \nl
 $\lj$ [pc] & 0.2-0.8 & 0.04-0.2 & 0.03-0.15 & 0.02-0.1  \nl
 ${\cal M}$ & 4 - 20 & 13 - 83 & 11 - 70 & 15 - 94  \nl
 $\ms$  & 1.3 - 2.2 & 2.2 - 4.2 & 1.5 - 2.9 & 1.8 - 3.4  \nl
$\vir$  & 1.2 - 0.2 & 0.5 - 0.05 & 0.2 - 0.02 & 0.15 - 0.02  \nl
\enddata
\label{Tablecas}
\end{deluxetable}
}
{\vskip 5.cm}


\begin{deluxetable}{lcccc}\,
\tablecolumns{5}
\tablewidth{0pc}
\tablecaption{Parameters defining the IMFs in eqn.~(\ref{IMFequ}), and the corresponding peak mass (in $dn/dM$, see eqn.(31)), for the different cases under study.
The value of the normalization constant $A_h$ corresponds to the mass integral equals to unity in eqn.(\ref{integmas}):
${\mathbb M}=1$.
}
\tablehead{
     & \colhead{MW}   & \colhead{Case 1}    & \colhead{Case 2} &  \colhead{Case 3} }
\startdata
$x$ & 1.35 & 1.35 & 1.6 & 1.6 \\
$m_0$  [$\msol$] & $2.0$ & $0.35$ & $0.35$ & 0.25 \\
$n_c$ & 11.0  & 14.0 & 11.0 & 14.0 \\
$\sigma$ & 0.579 & 0.607 & 0.531 & 0.558 \\
$m_c$  [$\msol$] & 0.18 & 0.025 & 0.032 & 0.018 \\
$A_h$  & 0.649 & 0.417 & 0.390 & 0.367\\\\
\hline\\
M$_{\rm peak}$ [$\msol$] & 0.03 & $4\times 10^{-3}$ & $5\times10^{-3}$  & $3\times 10^{-3}$ \\
\enddata
\label{TableIMF}
\end{deluxetable}
{\vskip 5.cm}

                          

\begin{deluxetable}{lcccccc}
\tablecolumns{6}
\tablewidth{0pc}
\tablecaption{Mass-to-light ratios ${\Upsilon_\star}$ in various bands at 10 Gyr obtained for the different IMFs given in Table 2 and for a Salpeter IMF.
 The last column gives the typical ratio ${\Upsilon_\star}/{\Upsilon}_{\star,MW}$.
The MW values are calculated with a \citet{Chabrier2005} IMF. Differences with a \citet{Chabrier2003} IMF are
 $\lesssim10\%$.}
\tablehead{
      &  B & V & R & I & K & ${\Upsilon}/{\Upsilon}_{\star,MW}$ }
\startdata
MW &  4.7 & 3.6 &  2.8 & 2.3 & 0.8 & \\
Salpeter   &  9.1 & 6.9 & 5.4 & 4.2 & 1.5 & $\sim$ 1.9 \\
Case 1   &  8.3 & 6.4 & 4.9 & 3.9 & 1.4 & $\sim$ 1.7\\
Case 2 &  10.3 & 7.9 & 6.1 & 4.8 & 1.7 & $\sim$ 2.1-2.2 \\
Case 3  &  11.1 & 8.5 & 6.5 & 5.1 & 1.8 & $\sim$ 2.3 \\
\enddata
\label{TableML}
\end{deluxetable}


\end{document}